%% file: o4vela.tex
\documentclass[twocolumn,times]{aastex631}

\usepackage{acronym}
\usepackage{amsmath}
\usepackage{xspace}
\usepackage{booktabs}
\usepackage{etoolbox}

\newtoggle{fullauthorlist}
\toggletrue{fullauthorlist}

\newtoggle{endauthorlist}
\toggletrue{endauthorlist}

\input{macros.tex}

\shorttitle{2024 Vela glitch GW constraints}
\shortauthors{LVK et al.}

\graphicspath{{./}{figures/}}

\begin{document}

\title{Constraints on gravitational waves from the 2024 Vela pulsar glitch}

\correspondingauthor{\lvkfull}
\email{lvc.publications@ligo.org}

\iftoggle{fullauthorlist}{
 \iftoggle{endauthorlist}{
  \suppressAffiliations
 }{}
 \input{LSC-Virgo-KAGRA-Authors-Feb-2025-aas.tex}
 \iftoggle{endauthorlist}{
  \collaboration{1}{\lvkfull}
  }{
  \collaboration{3000}{\lvkaff}
  }
}{
 \author{The LIGO Scientific Collaboration}\noaffiliation
 \author{the Virgo Collaboration}\noaffiliation
 \author{the KAGRA Collaboration}\noaffiliation
}

\input{em_authors.tex}
\iftoggle{endauthorlist}{
 \collaboration{20}{(See the end matter for the full list of authors and affiliations.)}
}{
 \nocollaboration{20}
}



\begin{abstract}
Among known neutron stars,
the Vela pulsar is one of the best targets for gravitational-wave searches.
It is also one of the most prolific in terms of glitches,
sudden frequency changes in a pulsar's rotation.
Such glitches could cause a variety of transient gravitational-wave signals.
Here we search for signals associated with a Vela glitch on \glitchdate
in data of the two LIGO detectors
from the fourth LIGO--Virgo--KAGRA observing run.
We search both for seconds-scale burst-like emission,
primarily from fundamental (f-)mode oscillations,
and for longer quasi-monochromatic transients
up to four months in duration,
primarily from quasi-static quadrupolar deformations.
We find no significant detection candidates,
but for the first time we set direct observational upper limits
on gravitational strain amplitude
that are stricter than what can be indirectly inferred from the overall glitch energy scale.
We discuss the short- and long-duration observational constraints
in the context of specific emission models.
These results demonstrate the potential of gravitational-wave probes of glitching pulsars
as detector sensitivity continues to improve.
\\
\textit{Draft revision: January 21, 2026}
\end{abstract}


\section{Introduction}
\label{sec:intro}

Pulsar glitches are sudden, enigmatic jumps
in a pulsar's spin frequency, which have been observed so far in over 200 pulsars~\citep{Espinoza2011,Yu2013,Basu2022}.
They generally involve an initial rapid change in frequency and frequency derivative
(on timescales of less than a minute)
followed by a more gradual
(on timescales of up to months)
relaxation towards the previous spin-down state
\citep{2022RPPh...85l6901A}.
While the common view is that glitches are linked to an exchange of angular momentum between an interior superfluid component and the crust of a \ns,
the quantitative details are still an open question.
Diverse mechanisms such as vortex avalanches, fluid instabilities and crust-quakes have been invoked as possible triggers \citep[see][for reviews]{1502.07062,2211.13885}.

Glitches are key tools for studying \ns interiors
\citep{10.1038/359616a0,astro-ph/0001245,1712.01547}
and could also trigger various types of \gw emission:
both transient signals linked to the initial jump,
and longer-lived signals linked to the relaxation
\citep{gr-qc/9711088,0809.4352,1008.0236,1104.1704,keer_jones_15,2015ApJ...807..132M,1605.08420,2003.12082,2401.05600,2007.05893,2204.12869,2403.09489,2406.00283,2024APh...15702921H}.
Studying these signals can constrain the \ns mass and radius and hence the nuclear matter equation of state.

Among the known pulsar population,
the Vela pulsar has some of the largest and most frequent glitches
\citep{1710.00952}.
It is also relatively close to Earth at a distance of \distvelapm\,pc~\citep{2003ApJ...596.1137D}.
For these reasons, it has been identified in several studies
\citep{2210.09907,2206.14515,2406.00283}
as a priority target to search for glitch-associated transient \gws.

Here, we present a set of searches for \gw emission
associated with the \glitchdate glitch of the Vela pulsar,
first observed \citep{atel16608,Zubieta:2025rud} by the
\IAR~\citep{2210.03770}
and subsequently confirmed by several other observatories
and precisely timed~\citep{atel16615} by the \MPRO of the University of Tasmania.
This glitch happened during the \lvk
\ohfour,
shortly after the two LIGO detectors \citep{1411.4547}
had restarted observations for the second part (O4b) of the run
and were joined also by Virgo \citep{1408.3978}.
The \elmag observations of the glitch are summarized in Section~\ref{sec:vela}
and the \gw data set is described in Section~\ref{sec:gwdata}.

To cover a range of glitch-related \gw emission scenarios (see \citealt{2024APh...15702921H} for a review),
our searches fall into two categories:
First, we search for transient \gw bursts that may be triggered by the initial rapid rise in frequency,
on dynamical timescales of milliseconds to seconds,
primarily from f-mode oscillations
at kHz frequencies,
in Section~\ref{sec:burst}.
Then, we cover longer-duration (up to four months)
quasi-monochromatic signals
at lower frequencies, coupled with the pulsar spin frequency and associated with the post-glitch relaxation,
in Section~\ref{sec:tcw}.
Even longer signals are well covered by standard \cw searches.
Details of the emission models are included in each of these sections.

Previous searches for \gws from pulsar glitches
have included an f-mode burst search
on the August 2006 Vela glitch
using initial LIGO data \citep{10.1103/PhysRevD.83.042001},
various all-sky burst searches \citep[latest O4a results:][]{2507.12282,2507.12374},
and long-duration searches
on O2 data \citep{1907.04717,2303.16720},
including the 2016 Vela glitch \citep{2018Natur.556..219P},
and O3 data \citep[][not including a Vela glitch]{2112.10990,2201.08785}.
However, matching the studies of detection prospects for the two cases by
\citet{2206.14515,2210.09907},
none of these searches had yet been able
to make a \gw detection
or to place physically constraining observational upper limits
below the energy scale indirectly
set by the glitch size.
Searches for \ns f-modes have also been performed on bursting magnetars,
with no detection yet either \citep[e.g.][]{2210.10931,2410.09151}.

The improved sensitivity of the LIGO \ohfour data \citep{2411.14607,2023PhRvX..13d1021G,2024Sci...385.1318J}
now gives us for the first time the opportunity to reach physically constraining results on \gws from a pulsar glitch.
We discuss their implications for pulsar glitch models
and \gw emission channels
in Section~\ref{sec:interpretation}
and present our conclusions in Section~\ref{sec:conclusions}.

\section{The Vela pulsar and its 2024 glitch}
\label{sec:vela}

The Vela Pulsar (PSR B0833$-$45/J0835$-$4510),
discovered by \citet{1968Natur.220..340L},
was the first pulsar known to glitch \citep{Radhakrishnan+1969c}.
It is young ($\sim 10^4\,\mathrm{yr}$)
and one of the most active pulsars in terms of glitching,
with 23 large glitches
(often considered as those with relative frequency changes
$\Delta \frot/\frot > 10^{-7}$,
\citealt{Espinoza2011,Yu2013,Basu2022})
reported so far --
approximately once every two to three years \citep{Reichley+1969, Dodson+2007}.
It also exhibits micro-glitches sporadically \citep{Cordes+1988,2021A&A...649C...2E,2023MNRAS.522.5469D, 2024A&A...689A.191Z}. 

During \ohfour, the Vela pulsar has been monitored with almost daily combined cadence by the
\IAR and \MPRO.

\subsection{IAR observations}

The IAR\footnote{\url{https://www.iar.unlp.edu.ar}} was founded in 1962 as a pioneer radio
observatory in South America.
Located near the city of La Plata, Buenos Aires, it is
located at
$34\degr 51\arcmin 57\arcsec.35$~S
$58\degr 08\arcmin 25\arcsec.04$~W.
It has two 30\,m diameter single-dish antennas,
aligned on a North--South direction, separated by
120\,m.
Observing around 1400\,MHz with 56\,MHz and 400\,MHz bandwidths simultaneously,
the data is obtained with a time resolution of $146~\mu$s.

A summary of IAR pulsar observation work and
future projects is given in \citet{2020A&A...633A..84G,2024RMxAC..56..134L}
and references therein.
The IAR has reported the observation of the last three large Vela glitches:
on 1 February 2019
with a relative frequency change of $2.7\times10^{-6}$
\citep{atel_vela2019,2020A&A...633A..84G},
on 22 July 2021
with relative change of $1.2\times10^{-6}$
\citep{2021ATel14806....1S},
and the \glitchdate glitch of interest for this paper
(counted as large Vela glitch \#23)
first reported in \citet{atel16608}
with $(\Delta\frot/\frot)=2.4\times10^{-6}$ 
and a more detailed analysis given in \citet{Zubieta:2025rud}.

For the data used in this paper,
we used the \texttt{PRESTO} \citep{2011ascl.soft07017R} package
to remove radio-frequency interferences (RFIs)
and fold the observations.
We cross-correlated the folded profile with a noise-free template profile in order to obtain the \toas.
We obtained the template with the tool \texttt{psrsmooth} in \texttt{PSRCHIVE} \citep{10.1071/AS04022} on a pulse profile with high signal-to-noise ratio (S/N).

\subsection{MPRO observations}

The University of Tasmania's \MPRO 26\,m radio telescope is located at Mt. Pleasant, near Hobart, Australia at
$42\degr 48\arcmin 12\arcsec.90$~S
$147\degr26\arcmin 25\arcsec.87$~E.
It has previously detected several Vela glitches,
including the real-time observation of the 2016 glitch
with associated magnetospheric changes \citep{2018Natur.556..219P},
which suggest a connection between the magnetosphere and the internal dynamics of the pulsar that probably provokes glitches.

The center frequency of the observations used here is 2250\,MHz with bandwidth 64\,MHz.
Processing was done using the software packages \texttt{DSPSR} \citep{10.1071/AS10021},
\texttt{TEMPO2} \citep{10.1111/j.1365-2966.2006.10302.x,astro-ph/0607664}, 
and \texttt{PSRCHIVE},
which perform phase-coherent de-dispersion using 16 frequency channels and 8192 pulse phase bins over the 89.3\,ms pulse period.
This gives a time resolution of $10.9\,\mu$s.

\subsection{Timing model}

We collected the \toas from both observatories between \mjd 60402.28 
and \mjd 60493.53 
and characterized the rotation of the Vela pulsar with \texttt{TEMPO2}
by fitting a Taylor series
\begin{equation}\label{eq:timing-model}
    \phi(t) =
    \frot(t-\Tref)
    + \frac{1}{2}\fdotrot(t-\Tref)^2
    + \dots \,.
\end{equation}
The pulsar's rotation frequency is $\frot$, its derivatives are $\fdotrot$, $\fddotrot$, etc., and $\Tref$ is the reference epoch. 

To characterize the glitch, we added a glitch-induced change in the rotational phase \citep{astro-ph/0607664,Yu2013}
to this model:
\begin{multline}\label{eq:glitch-model}
    \phi_\mathrm{g}(t)
    = \Delta \phi
    + \Delta \frot (t-\Tgl)
    + \frac{1}{2} \Delta \fdotrot (t-\Tgl)^2
    \\ 
    + \sum_i \left[1-\exp{\left(-\frac{t-\Tgl}{\tau^{i}_\mathrm{d}}\right)} \right]
    \Delta \frot^{d,i}\, \tau_\mathrm{d}^{i} \,.
\end{multline}
Here, $\Tgl$ is the epoch of the glitch,
with its uncertainty counteracted by $\Delta \phi$,
while
$\Delta\frot$ and $\Delta\fdotrot$ are the sudden changes in $\frot$ and $\fdotrot$ at $\Tgl$.
In addition, $\Delta\frot^{d,i}$ are the temporary increases (recovery terms) in the rotational frequency of the pulsar that decay exponentially on a timescale $\tau_\mathrm{d}^{i}$ days.

\begin{figure}[ht!]
\includegraphics[width=\columnwidth]{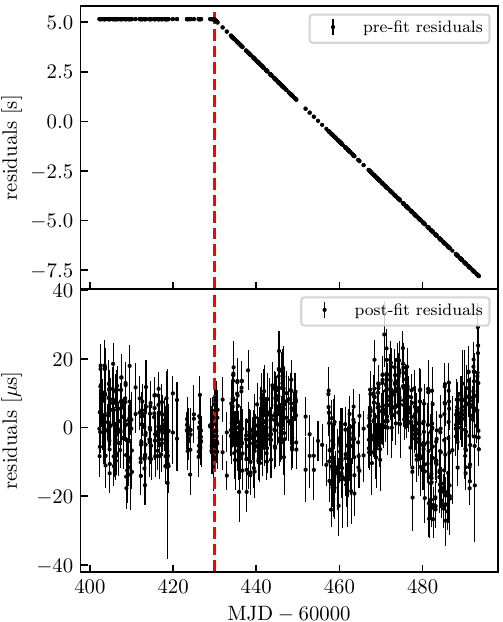}
\caption{ 
\label{fig:Vela-residuals}
Top panel: Vela pulsar radio timing residuals before fitting for the \glitchdate glitch.
Bottom panel: Timing residuals after fitting for the glitch with the parameters shown in Table~\ref{tab:Vela_glitch}.}
\end{figure}

The \glitchdate Vela glitch is evident from the pre-fit residuals in Figure~\ref{fig:Vela-residuals}.
We characterized the pre-glitch timing model by fitting Equation~\eqref{eq:timing-model} to \toas from one month of observations before the glitch.
By combining the \IAR and \MPRO data, we found preference for three transient increase terms in the frequency of the pulsar in addition to the permanent changes $\Delta\frot$ and $\Delta\fdotrot$.
The post-fit residuals can be seen in the bottom panel of Figure~\ref{fig:Vela-residuals}. We then extended the data span to \mjd 60357--60551, which is the full range of interest for the GW search here. With this data span, we fitted $\frot$, $\fdotrot$, $\fddotrot$ and $\fdddotrot$ keeping the glitch parameters frozen. The full timing solution covering the extended range of observations
is shown in Table~\ref{tab:Vela_glitch}.
This is valid within the listed START and FINISH dates,
and hence the listed $\fddotrot$ should not be interpreted as giving the long term braking index~\citep[see][]{2305.09079}.

\begin{deluxetable}{lr}
\tablecaption{
 \label{tab:Vela_glitch}
 Parameters of the Vela pulsar and its 2024 glitch.
}
\tablehead{ \colhead{parameter} & \colhead{value}}
\startdata
R.A. & 08:35:20.61993 $\pm\;0.00002$ \\ 
DEC. & -45:10:33.7232 $\pm\;0.0003$ \\ 
$d$ [pc] & \distvelapm$^{(a)}$\\
$\Tref$ [d] (MJD) & 60408 \\ 
$\Tref$ [s] (GPS) & 1396569549 \\
START (MJD) & 60357 \\
FINISH (MJD) & 60551 \\
$\frot$ [Hz] & $(11.182888898688 \pm 0.000000000002)$ \\
$\fdotrot$ [Hz\,s$^{-1}$] & $(-1.5541536 \pm 0.000001) \times 10^{-11}$ \\
$\fddotrot$ [Hz\,s$^{-2}$] & $(3.82\pm 0.03)\times 10^{-22}$\\
$\fdddotrot$ [Hz\,s$^{-3}$] &  $(1.283\pm0.07) \times 10^{-28}$\\
$\Tgl$ [d] (MJD) & $60429.86975 \pm 0.00139$ \\
$\Tgl$ [s] (GPS) & $1398459095.216 \pm 120.096$ \\
$\Delta\phi$ & $-0.00034 \pm 0.00017$ \\
$\Delta\frot$ [Hz] & $(2.65854 \pm 0.00005) \times 10^{-5}$\\
$\Delta\fdotrot$ [Hz\,s$^{-1}$] & $(-1.0298 \pm 0.0009)\times 10^{-13}$ \\
$\Delta\fddotrot$ [Hz\,s$^{-2}$] & --- \\
$\tau_d^{1}$ [d]& $15.1\pm0.1$ \\
$\Delta \frot^{d,1}$ [Hz] & $(1.501\pm0.004) \times 10^{-7}$ \\
$\tau_d^{2}$ [d] & $2.45\pm0.06$ \\
$\Delta \frot^{d,2}$ [Hz] & $(1.24\pm0.03) \times 10^{-7}$ \\
$\tau_d^{3}$ [d] & $0.39\pm0.03$ \\
$\Delta \frot^{d,3}$ [Hz] &  $(1.8\pm0.1) \times 10^{-7}$ \\
\enddata
\tablecomments{
 (a) \citet{2003ApJ...596.1137D} \\
 All other values were derived from
 \IAR and \MPRO observations,
 as described in Section~\ref{sec:vela}.
 \\
 Reference time and glitch epoch
 (at the solar system barycenter)
 are given in both
 MJD (terrestrial time standard),
 as is conventional in radio astronomy,
 and in GPS seconds (UTC standard)
 for \lvk data analysis.
 $\Tgl$ is given to the precision ingested by burst analyses.
}
\end{deluxetable}

\section{GW data} \label{sec:gwdata}

The \ohfour run of the advanced laser-interferometric \gw detectors
began on 27 May 2023
and, after a commissioning break
with several interventions to further improve sensitivity,
its second part O4b ran from 10 April 2024 to 28 January 2025.
In O4b,
the two LIGO detectors \citep{1411.4547}
in Hanford (Washington, US) and Livingston (Louisiana, US)
and the Virgo detector \citep{1408.3978} in Cascina (Italy)
participated.
KAGRA \citep{2005.05574} in Japan rejoined \ohfour later.
The upgrades to LIGO for \ohfour
allowed it to reach unprecedented \gw strain sensitivity
\citep{2411.14607,2023PhRvX..13d1021G,2024Sci...385.1318J}.

\begin{figure*}[ht!]
\includegraphics[width=\linewidth]{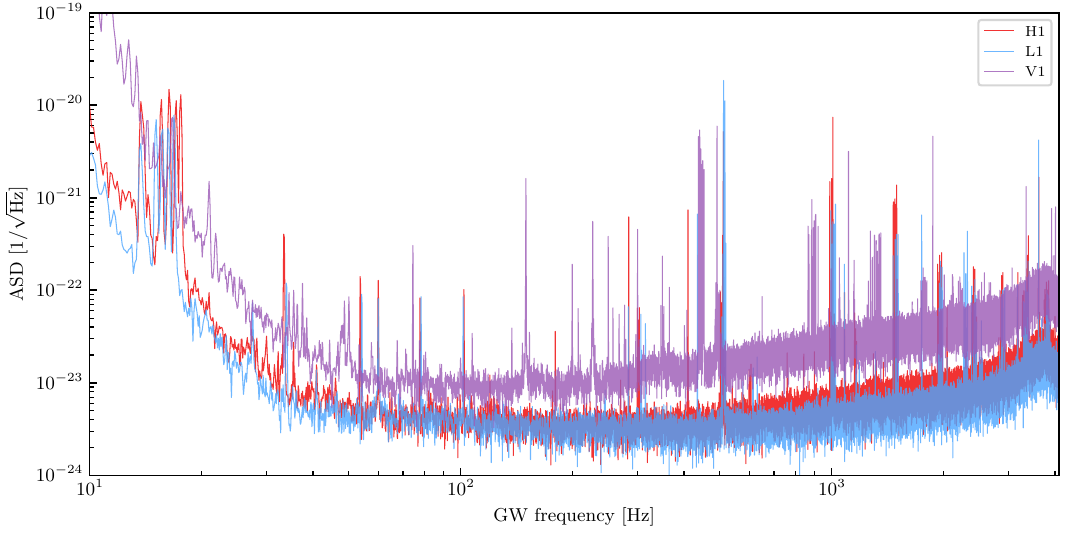} \\
\includegraphics[width=\columnwidth]{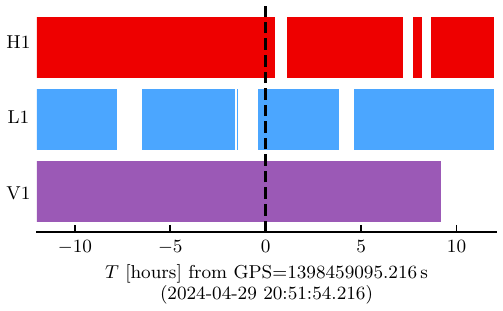}
\includegraphics[width=\columnwidth]{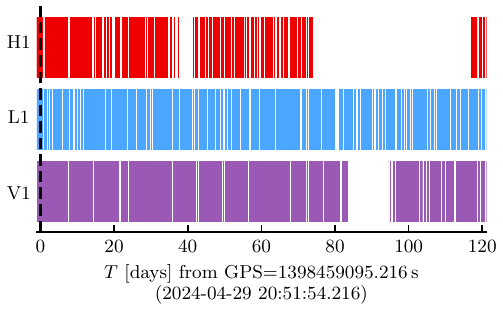}
\caption{ 
\label{fig:gw_data}
Top panel: \gw detector sensitivities within $\Tgl\pm20$\,s around the Vela glitch on \glitchdate,
in terms of their \asds
(computed with the Welch method in \texttt{gwpy}, \citealt{gwpy}).
Bottom left panel: GW data availability over 24\,h around the glitch,
as used for the burst searches in Section~\ref{sec:burst}.
Bottom right panel: LIGO data availability for the four following months,
as used in the long-duration searches in Section~\ref{sec:tcw}.
Virgo data is not used in either type of searches.
The glitch time is highlighted in the two bottom panels with a vertical dashed line.
}
\end{figure*}

At the time of the Vela glitch on \glitchdate,
the LIGO Hanford (H1),
LIGO Livingston (L1)
and Virgo (V1)
detectors were all operating in a stable state;
see Figure~\ref{fig:gw_data}.
The burst-type searches in Section~\ref{sec:burst}
use data within two days of the glitch time.
The coincident data taking of the two LIGO detectors with similar sensitivity across most of the frequency band was essential,
as a single-detector search of this type
loses an order of magnitude in sensitivity~\citep[e.g.,][]{ball_joint_implications}
due to missing suppression capability towards instrumental glitches
(not to be confused with pulsar glitches).

The longer-duration searches in Section~\ref{sec:tcw}
cover a maximum extent
starting one day before the glitch epoch
and ending on 28 August 2024.
For part of this time stretch,
only L1 data are available,
as H1 had to be taken offline
to diagnose and repair a problem in the output optics chain
from 13 July to 24 August,
and V1 went into a downtime for limiting noise factors
from 22 July to 2 August.
The analyzable LIGO data segments \citep{T2400058} over this extended timescale
are also shown in Figure~\ref{fig:gw_data}.

Due to the large difference in sensitivity,
in terms of the detectors' \asds,
Virgo data is not used in this paper.
From LIGO, we use data from the \texttt{GDS-CALIB\_STRAIN\_CLEAN} frame channel,
corresponding to the standard online calibration \citep{viets2018reconstructing, Sun:2020wke, Sun:2021qcg,Wade:2025tgt}
with some noise subtraction applied \citep{T2100058,1911.09083,O4gwosc}.
For several searches in Section~\ref{sec:tcw},
additional time-domain removal of large noise excursions
was performed \citep{Astone:2005fj,gating},
before creating Fourier-domain
data products as described later for each analysis method.
Data quality has been investigated with methods similar
to those described for O4a in \citet{2409.02831}.

\section{Searches for GW bursts}
\label{sec:burst}

\begin{figure}[t!]
\includegraphics[width=\columnwidth]{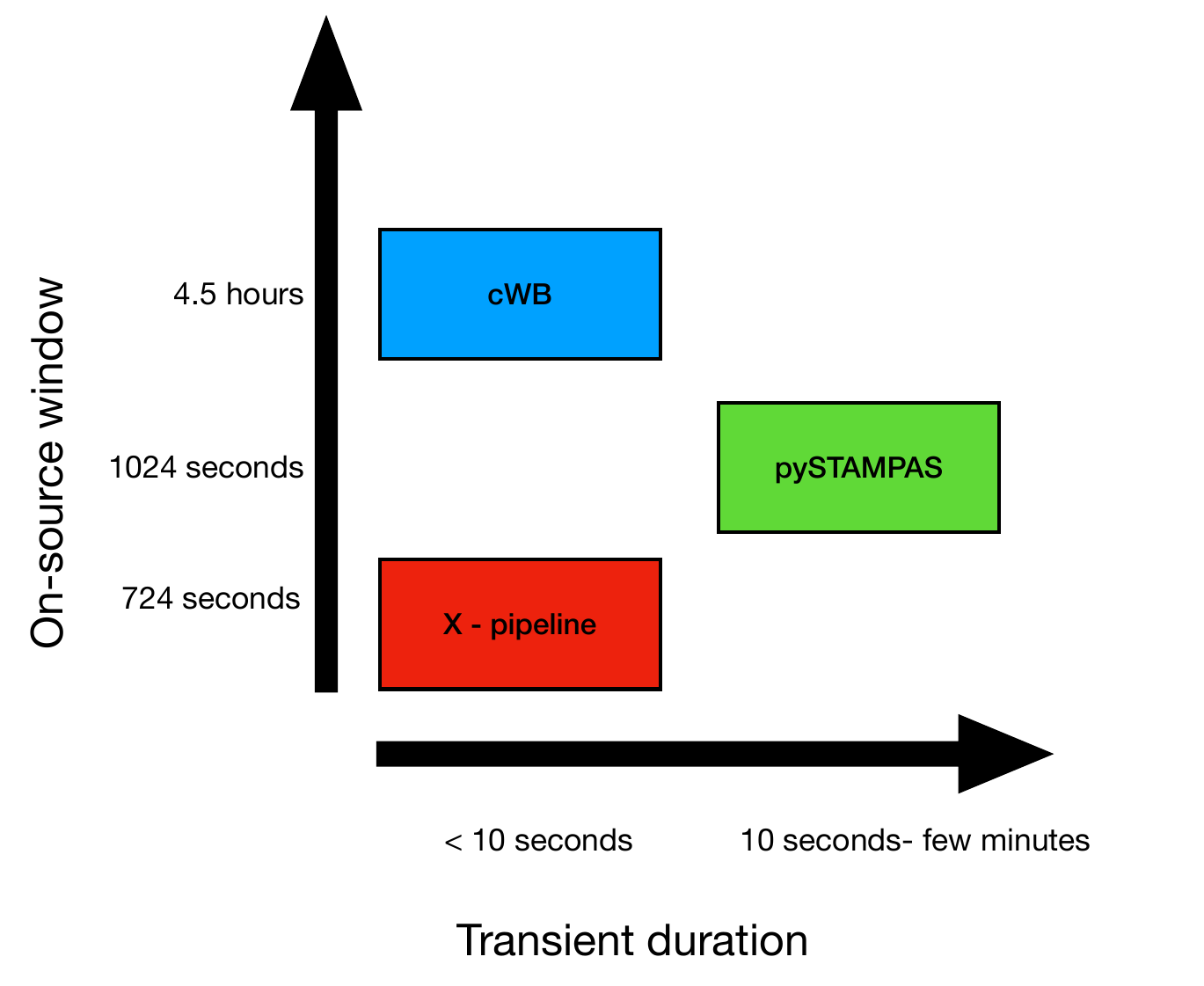}
\caption{ 
\label{fig:burstsoverview}
This schematic summarizes the burst-like transient searches performed for this paper as a function of the duration of anticipated signals and the on-source windows used by each.
The on-source window time for \xpipe is 1$\sigma$ uncertainty around the estimated glitch arrival time using the timing model, the on-source window for \cwb is the broadest window in radio observation within which the glitch occurred.
The choice for \pystampas is based on analysis requirements and longer duration of the signal.   
}
\end{figure}

In the context of this paper, we refer to any signals as ``\gw bursts'' that last from milliseconds up to, at most, a few minutes.
The main physical emission channel considered are \ns f-modes excited by the glitch and damping away again on the signal timescale \citep{gr-qc/9711088}.
However, all three search methods that we use here are unmodeled or very weakly modeled, i.e., they do not rely on matched filtering.
Thus, they are also sensitive to other emission mechanisms in the same time and frequency ranges.
The allowed signal durations and the on-source windows (time range in which the signal might start) for the three methods  are summarized in Figure~\ref{fig:burstsoverview}.
Here, the glitch time is the GPS time at the detector.

These, or similar, methods have been routinely used for all-sky \gw burst searches in \lvk data \citep[e.g.,][]{2107.03701,2107.13796,2507.12282,2507.12374},
which would have also been sensitive to signals from pulsar glitches with sufficiently high amplitudes \citep{2206.14515},
and also for analyzing bursting magnetars \citep[e.g.][]{2210.10931,2410.09151}
However, the last targeted search for \gw bursts from a pulsar glitch dates back to the era of the initial LIGO detectors,
when the 2006 Vela glitch was analyzed \citep{10.1103/PhysRevD.83.042001} using the two Hanford detectors available at the time.
With the two Advanced LIGO detectors and improved analysis methods, we now obtain orders of magnitude better results than that study.

In this section,
after briefly discussing the f-mode scenario,
we introduce the three pipelines used
and summarize their results.
They cover different time and frequency ranges
to increase robustness to different emission scenarios.
For each search,
after finding no significant candidates,
we conduct injections of simulated signals into the \gw data to estimate search sensitivity and provide upper limits; this is discussed in Section~\ref{sec:bursts_sensitvity}.  
Detailed interpretation is left to Section~\ref{sec:interpretation}.

\subsection{Emission Models}
\label{sec:burst-emission}

Glitching pulsars can emit short-duration \gw signals
via mechanisms such as
an avalanche of unpinning superfluid vortices \citep[see, e.g.,][and references therein]{2012MNRAS.423.2058W,2023PhRvD.107j2001D} or oscillation modes of the NS \citep{gr-qc/9711088,keer_jones_15,2204.12869,2403.09489}.

The situation is complicated, as \nss can support a variety of oscillations modes, classified according to the dominant restoring force, e.g.\ pressure, buoyancy, or rotation \citep{1999LRR.....2....2K}. 
We mostly focus on the fundamental (f-)modes, which are good candidates for our searches for several reasons:
They are well understood, with frequencies depending only on the gross features of the \ns equation of state, typically in the kHz range accessible by \lvk detectors.
They are also extremely efficient \gw emitters, in the sense that almost all energy deposited in the mode by the glitch is emitted in quadrupolar \gws, with very little energy loss to viscosity \citep{PhysRevD.40.3194,Finn:1992wt,2003.12082}.

Regardless of the precise emission mechanism, we can define a \emph{characteristic energy scale} as the apparent change in kinetic energy of the \ns caused by the glitch,
\begin{equation}
\label{eq:E_characteristic}
\Delta \Echar \equiv 4\pi^2 I \frot \Delta \frot \,,
\end{equation}
where $I$ is the \ns moment of inertia.
Inserting values for $\frot$ and $\Delta \frot$ from Table~\ref{tab:Vela_glitch} and a fiducial $I = 10^{45} \, \text{g cm}^2$,
we obtain $\Delta \Echar \approx 1.15 \times 10^{43}$ erg.

The relation of this energy to actual radiated \gw energies is highly model-dependent; see \citet{2406.00283} for a summary of the literature.
In particular, the energy transferred to the f-mode could be much smaller than $\Delta \Echar$
if the timescale associated with the glitch trigger (possibly crust cracking, or vortex unpinning)
is much longer than the ${\cal O} \sim 1$\,ms timescale characteristic of an f-mode oscillation, as discussed in the context of magnetar flares in \citet{levin:2011}.
It is also possible that other modes, with oscillation periods closer to the glitch trigger timescale, might be more efficiently excited, e.g., the torsional crustal oscillations considered in the context of Vela glitches in \citet{bransgrove:2020};
such oscillations are relatively weak \gw emitters.  

With these caveats in mind, we will nevertheless choose f-modes as the reference model for evaluating the performance of our (unmodeled) searches, and use $\Delta \Echar$ as a useful energy scale to compare our results against.
For a signal starting at $t_0$,
the f-mode signal can be written as \citep{10.1103/PhysRevD.83.042001}
\begin{subequations}
  \label{bursts:ringdown}
 \begin{align}
   h_+(t)      = A_+ \cos\left(2 \pi \fgw (t-t_0) + \phi_0\right) e^{-(t - t_0) / \tau} \,, \\
   h_\times(t) = A_\times \sin\left(2 \pi \fgw (t-t_0) + \phi_0\right) e^{-(t - t_0) / \tau} \,. 
 \end{align}
\end{subequations}
The amplitudes $A_+$ and $A_\times$ depend on the \ns inclination angle $\iota$
as $1 + \cos^2(\iota)$
and $2 \cos(\iota)$.
$\tau$ is the damping time,
and $\phi_0$ is the initial phase.
The frequency and damping time depend on the equation of state of \ns matter,
with ranges of $\sim[1,4]$\,kHz and $\lesssim1$\,s
for most realistic models
\citep[see, e.g.,][]{gr-qc/9711088,2403.09489}.
These ranges are covered by the searches described below.

The actual emission from a glitching \ns can be more complex than the simplistic assumptions made here.
To mitigate this, we have employed three pipelines that do not depend explicitly on the morphology of the incoming \gw signal.
The damped sinusoid model of Equation~\eqref{bursts:ringdown} is then used for sensitivity measures and computing upper limits. 

\subsection{coherent WaveBurst (\texorpdfstring{\cwb}{cWB})}

\cwb \citep{Klimenko:2015ypf,Drago:2020kic,Martini:2025ewk} is a \gw search and reconstruction algorithm based on maximum-likelihood statistics that uses the excess signal power in the detector network and does not rely on a precise signal model of the incoming \gws. 
The time series strain data are transformed into a multi-resolution time-frequency map using the Wilson--Daubechies--Meyer (WDM) wavelet transform \citep{Necula:2012zz}, then excess power clusters of pixels are identified for each detector in the network, and a likelihood is computed for these selected pixels as a function of sky direction.
\cwb is used routinely for the search and reconstruction of \gw transients in \lvk data \citep{2107.03701,KAGRA:2021vkt}.
The version of the pipeline used here is the same as in the O4a all-sky short-duration search \citep{2507.12374}.
\cwb is sensitive to signals expected from NS glitches, such as f-modes \citep{2206.14515},
and was recently used for all-sky burst searches \citep{2107.03701,KAGRA:2021vkt,2507.12282,2507.12374}.

The \cwb analysis searches for short-duration transient signals (less than 10 seconds) in the frequency range of 512\,Hz to 4000\,Hz.
It covers a time window of 4.5 hours around the estimated glitch time (at Earth, rounded to the nearest integer, corresponding to 1398458994 GPS seconds), addressing the broadest possible uncertainty from radio observations. 
The analysis is restricted to sky directions within 1 square degree centered around the Vela pulsar.
For the 4.5\,hours of on-source time window we perform time slides to estimate the  background~\citep{Was:2009vh}.
The total background generated is 6.6\,years, allowing false-alarm probability estimates at the level of $7.5 \times 10^{-5}$. 
The search does not find any triggers in the on-source time window above the thresholds applied within the pipeline.

\subsection{\texorpdfstring{\pystampas}{PySTAMPAS}}

\pystampas \citep{2021PhRvD.104j2005M} is an unmodeled coherent search pipeline based on the cross-correlation of multi-resolution time-frequency maps and targeting long duration \gw transient signals.
The analysis uses the same version of the pipeline as in the O4a all-sky long-duration bursts search \citep{2507.12282}.
The pipeline splits data into 512\,s segments and generates signal-to-noise-ratio spectrograms for each detector using Short-Time Fourier Transforms (STFTs) over short windows (0.5, 1, 2, and 4 seconds, each Hann‑windowed with 50\% overlap) to form multi‑resolution spectrograms covering 512--4020\,Hz, with the \psd in each frequency bin estimated as the median of the squared STFT magnitudes.
A seed‑based clustering algorithm then identifies candidate triggers, which are cross‑correlated with the corresponding spectrogram from the second detector to compute a coherent detection statistic.
The search is performed considering the Vela sky position given in Table~\ref{tab:Vela_glitch}.
The on-source window is $\pm$ 512\,s long centered at the glitch time 1398458994, to ensure covering the full signal power also for longer-duration signals of 50--100 seconds. 
We consider two days of data around the glitch time excluding the on-source window to estimate the background distribution of noise triggers by 25 time-shifts of the time-frequency maps by more than 1000\,s to effectively simulate 31 days of background data.
This procedure establishes a $3\sigma$\, detection threshold corresponding to a false-alarm rate (FAR) of $3.3\times10^{-6}$\,Hz.
The search does not find any event in the on-source time window below this FAR, the loudest event has FAR more than an order of magnitude higher corresponding to p-value of around $0.6$ which is compatible with Gaussian noise fluctuations. 

\subsection{\texorpdfstring{\xpipe}{X-pipeline}}
The \xpipe is an un-modeled coherent search pipeline where the data from each detector is coherently combined in the multi-resolution time-frequency domain.
The brightest 1\% of pixels are selected along with nearby pixels that form a cluster and then assigned likelihood-based ranking statistics \citep{Sutton:2009gi,Was:2012zq}.
Spurious noise triggers are rejected by applying vetoes to the coherent and incoherent power across the detector network. 

The search performed for this paper was done on an on-source time window of $\pm 362$ seconds centered at GPS time 1398458994, representing $1\sigma$ uncertainty in the glitch time estimate.
The frequency band of the analysis is 200\,Hz to 4000\,Hz.
The off-source time window used for background computation is $\pm 1.5$\,hours around the glitch time.
The loudest event found in the analysis has p-value of 0.8 which is compatible with Gaussian noise fluctuations.

\subsection{Search Sensitivity} 
\label{sec:bursts_sensitvity}

While the burst searches are independent of a specific \gw waveform,
to assess their sensitivity
we inject simulated damped sinusoid waveforms,
as a simple stand-in model for emission from $f$-modes,
into the detector data around the time of the glitch.
These injections are fixed to the sky direction and inclination of the Vela pulsar.
Since the \ns equation of state and the precise glitch mechanism are not known, we also use a range of central frequencies and damping times for the damped sinusoids.
We list the injection sets used for each pipeline in Table~\ref{tab:Bursts_Inj}.
These cover different timescales due to the different search ranges as previously summarized in Figure~\ref{fig:burstsoverview}
and,
for practical reasons,
somewhat different but overlapping frequency ranges.

We then vary the amplitude of each damped sinusoid to obtain an efficiency curve for each pipeline, ranging from 0 to 100\%.
We fix a detection threshold of $3\sigma$ for each pipeline,
which corresponds to inverse false-alarm rates of 0.2 years, 0.008 years and 0.009 years for \cwb, \xpipe and \pystampas, respectively due to differences in the on-source time window.
The upper limits apply only to signals during coincident data availability from the LIGO detectors, similar to choices made in previous works \citep[e.g., the search for \gws from SN2023ixf][]{LIGOScientific:2024jxh}.

\begin{table}[t!]
\centering
\caption{
\label{tab:Bursts_Inj}
Parameters of the damped sinusoid injection sets used by each pipeline}
\begin{tabular}{@{}lll@{}}
\toprule
\textbf{Pipeline} & \textbf{$\fgw$(Hz)} & \textbf{$\tau$(s)} \\
\midrule
\cwb & \begin{tabular}[t]{@{}l@{}}650, 1100, \\2100, 3900\end{tabular} & 0.01, 0.1, 0.5 \\
\xpipe & \begin{tabular}[t]{@{}l@{}}290, 650, 1100, \\2100, 3900\end{tabular} & 0.01, 0.1, 0.5 \\
\pystampas & \begin{tabular}[t]{@{}l@{}}800, 1050, 1300, \\ 
1600, 2000, 2200, \\ 
2500, 2750, 3000, 3250\end{tabular} & 10 \\
\bottomrule
\end{tabular}
\end{table}

\begin{figure}[ht!]
\includegraphics[width=\columnwidth]{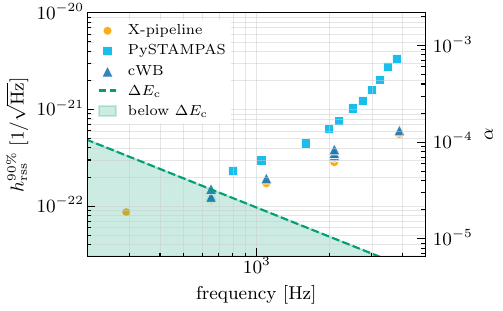}
\caption{ 
\label{fig:bursts-sensitivity}
$\hrss^{90\%}$ at a detection threshold of $3 \sigma$ (p-value $10^{-3}$)
for the three \gw burst searches (circle, square and triangle markers),
shown against the frequency of damped sinusoids signals.
For \cwb and \xpipe, multiple markers at fixed frequencies denote different damping times (see Table~\ref{tab:Bursts_Inj}), which do not effect the sensitivity drastically.
Strains corresponding to the characteristic energy $\Delta\Echar$ of the glitch
from Equation~\eqref{eq:E_characteristic}
are shown as the dashed green line
and \gw emission would be consistent with it in the shaded region below.
The right-hand vertical axis shows the r-mode amplitude $\alpha$ that would correspond to a given $\hrss$ level.
See Section~\ref{sec:interpret_bursts} for interpretation of these results in terms of $\Delta\Echar$ and $\alpha$.
}
\end{figure}

We present the results in terms of the quantity 
\begin{equation}
\label{eq:h_rss_definition}
\hrss = \sqrt{\int_{-\infty}^{\infty} \left(h_{+}^{2}(t) + h_{\times}^2(t)\right) dt } \,
\end{equation}
where $h_+(t)$ and $h_{\times}(t)$ are the amplitudes for the two \gw polarization.
The Vela pulsar's inclination angle dictates the relative content of $h_+$ and $h_{\times}$ of the waveform.
We perform a set of injections over a range of amplitudes of $[10^{-23},10^{-21}]$ and recover them with each of the search pipelines,
then measure the $\hrss$ needed to achieve the detection efficiency 90\%.
We show this $\hrss^{90\%}$,
at a $3\sigma$ detection threshold
and as a function of \gw frequency,
for all pipelines and various damping times in Figure~\ref{fig:bursts-sensitivity}.

The results are different due to both methodological aspects and the different parameter spaces,
as summarized in Figure~\ref{fig:burstsoverview}.
\pystampas probes longer-duration signals,
which are more difficult to detect with un-modeled methods.
The difference between \xpipe and \cwb can be in part explained by \xpipe's tighter on-source window.
Even with the same false-alarm rate threshold between the pipelines, there would still be variations due to the antennae pattern functions and noise properties changing over the different time windows, hence a fair comparison cannot be made and is also out-of-scope of this work.

In the same figure we have also shown the $\hrss$ corresponding to \gws with the full characteristic glitch energy
from Equation~\eqref{eq:E_characteristic};
see Section~\ref{sec:interpret_bursts} for its derivation and interpretation.
The \xpipe and \cwb results
compare more favorably with the available energy budget
below 1\,kHz
than at higher frequencies,
while for \pystampas only injections at higher frequencies were done.

In the O4a all-sky burst search~\citep{2507.12374},
the minimum relative glitch size for Vela-like pulsars to be detected was around $\Delta\frot/\frot \approx 2 \times 10^{-5}$, which is an order of magnitude larger than the $\Delta f_\glitch$ of this glitch.
But the search sensitivity in that paper was computed with injections above 2000\,Hz, which along with all-sky and all-time averaging makes those results not directly comparable to the ones in this paper.

\section{Searches for quasi-monochromatic long transient GWs}
\label{sec:tcw}

Quasi-monochromatic long-duration \gw signals
at frequencies coupled to the pulsar's rotation
may be triggered by glitches \citep{1104.1704}
and can be searched for with methods derived from those for \cws;
see \citet{2206.06447} for a review.
The first such searches were done
on O2 data \citep{1907.04717},
including the 2016 Vela glitch \citep{2018Natur.556..219P}
and a glitch of the Crab pulsar,
and on O3 data \citep{2112.10990,2201.08785},
covering six glitching pulsars but with no Vela glitch during that run.
Detection prospects for this type of \gw signals were studied in \citet{2210.09907},
indicating that Vela glitches from O4 on would allow for physical constraints
below the indirect energy estimate previously introduced by \citet{1104.1704}, equivalent to our Equation~\eqref{eq:E_characteristic}.
\citet{2406.00283} also identified Vela as a priority search target.

In this section,
we first discuss emission models for this type of signal.
We then present several analyses that differ in the strictness of their model assumptions:
two fully-coherent and two semi-coherent search methods,
each with different allowances for the degree of mismatch between the \gw and \elmag phase evolution
and varying flexibility in signal start-times and durations.
The motivation is to probe as deep as possible into the noise floor for the strictest model assumptions,
while also maintaining chances to detect sufficiently strong signals
with different parameters or time evolution.

As there are no significant detection candidates from either search,
we report upper limits on \gw strain from all of them,
which are compared in Section~\ref{sec:tcw-uls}
and Figure~\ref{fig:cw_uls}.
More detailed results from each method
are contained in the appendices.
Detailed interpretation is left to Section~\ref{sec:interpretation}.

\subsection{Emission Models}
\label{sec:tcw-emission}

Models for transient \cw-like emission following a pulsar glitch,
as summarized by \citet{2024APh...15702921H} and \citet{2406.00283},
can be divided into two classes:
\gw emission caused by motions of the fluid core,
or by quadrupolar deformations supported by the solid outer crust. 

For fluid motion in the core, the leading model involves \emph{Ekman pumping},
where the sudden spin-up of the crust induces a meridional flow in the interior fluid \citep{0809.4352}, which may be non-axisymmetric.
Calculations by \citet{1008.0236} and \citet{1605.08420}
found emission via both the mass and current quadrupoles
and at both the $\frot$ and $2\frot$ harmonics.
The \gw amplitude depends on the \ns equation of state via the matter compressibility.

For asymmetries in the crust,
the key idea is that,
for any combination of elastic fracture and superfluid unpinning causing the glitch,
the process happens in a non-axisymmetric way,
so that a large-scale (quadrupolar) deformation is produced in the solid outer crust,
which then relaxes away on some (poorly constrained) viscous timescale of the crust.
This relaxation may leave an imprint in the post-glitch spin evolution,
as exploited in the model of \citet{2007.05893},
which attributes the exponentially-decaying excess spin-down seen after most glitches to the formation of a transient mountain,
causing \gw emission that provides the increased spin-down torque.

We can again use the characteristic glitch energy scale of Equation~\eqref{eq:E_characteristic} to describe such transient-\cw emission,
but inserting
a dimensionless parameter $Q$ describing how much of this energy is radiated in \gws:
\begin{equation}
\label{eq:E_tCW_with_Q}
    \Delta E_{\rm c, \, tCW} 
    = 4\pi^2 I \frot \Delta \frot Q \,.
\end{equation}

In the \citet{2007.05893} transient mountain model,
only the part of $\Delta \frot$ that decays away in the post-glitch relaxation contributes to \gw emission, at $2\frot$.
In this case,
the fraction of energy corresponds to the \emph{healing parameter}
\begin{equation}
 \label{eq:healingQ}
 Q = \frac{\Delta \frot^d}{\Delta \frot}
\end{equation}
as usually defined in pulsar timing
\citep{2022RPPh...85l6901A}.
For the 2024 Vela glitch,
$\Delta \frot^d$ can be taken as the sum of the three decay components identified in radio pulsar timing
(Section~\ref{sec:vela}).
Using the parameters in Table~\ref{tab:Vela_glitch}, we find $Q \approx \Qvalue$, i.e., only about $1.7\%$ of the characteristic glitch energy would be radiated in \gws in this scenario.

More energetic emission mechanisms
with $Q \sim 1$ have been speculated upon (see section II.C. of \citealt{1104.1704} and Table 1 of \citealt{2406.00283}).
However, these scenarios lack the detailed emission mechanisms of the Ekman pumping or Yim \& Jones decaying mountain models, reflecting our uncertainty in the nature of transient \gw emission. 

Furthermore, if the \ns is excited into \emph{free precession} following the glitch, a \cw-like signal would be emitted, decaying on a timescale determined by the strength of dissipative processes within the star \citep{PhysRevD.20.351, 2002MNRAS.331..203J}.
The signal would contain harmonics at (or close) to both $\frot$ and $2\frot$, with the lower harmonic dominating in the case of a small angular misalignment between the \ns symmetry axis and its (fixed) angular momentum vector.
Such precessional motion would also leave an imprint on the radio pulsar signal, modulating the frequency and pulse profile \citep{2001MNRAS.324..811J}.
No such modulations have been reported, although they may occur with too small an amplitude and/or on too long a timescale to be apparent. 

Given the wide range of long-transient \gw emission mechanisms and strengths found in the literature, we carried out a number of relatively agnostic searches, as described below.
All allow for a wide range of signal durations,
and some also for emission at both the $\frot$ and $2\frot$ harmonics.
Additionally,
we consider \gw amplitude evolution profiles
that are either constant, i.e., set by a rectangular window function
\begin{eqnarray}
\label{eq:cw_windwo_rect}
h(t; \lambda, \mathcal{A}, \mathcal{T}) = \begin{cases} h(t; \lambda, \mathcal{A}) & \text{if~} t_0 \le t < t_0+\tau \,; \\
0 & \text{otherwise},
\end{cases}
\end{eqnarray}
a single exponentially decaying window function,
\begin{eqnarray}
\label{eq:cw_windwo_exp}
h(t; \lambda, \mathcal{A}, \mathcal{T}) = \begin{cases} e^{-(t - t_0) / \tau} h(t; \lambda, \mathcal{A}) & \text{if~} t \ge t_0 \,; \\
0 & \text{otherwise},
\end{cases}
\end{eqnarray}
or a combination of three exponentials, following Equation~\eqref{eq:glitch-model}.
Here, $h(t; \lambda, \mathcal{A})$ would be a standard persistent \cw signal \citep{gr-qc/9804014,2206.06447}
with two polarization components,
described by amplitude parameters $\mathcal{A}$
(including a constant characteristic strain amplitude $h_0$,
inclination $\iota$,
polarization angle $\psi$,
and an initial phase)
and phase-evolution parameters $\lambda$ (frequency, frequency derivatives, and sky position);
and $\mathcal{T}=\{t_0,\tau\}$
are the transient window parameters:
$t_0$ is the start of the signal window
and $\tau$ is the rectangular window length
or exponential decay constant.
In the following, we also define for convenience
$h_0(t) \equiv e^{-(t - t_0) / \tau} h_0$
in the exponentially decaying case,
with $h_0(t=t_0)=h_0$.

All the searches and upper limits presented here
are set up to correctly take into account the gaps in \gw data
(as seen in Figure~\ref{fig:gw_data}),
as is standard for \cw search methods.

\subsection{Time-domain Bayesian search (\cwinpy)}
\label{sec:tcw-cwinpy}

The time-domain Bayesian method \citep{10.1103/PhysRevD.72.102002, 10.48550/arXiv.1705.08978} has been widely used for targeted \cw searches \citep[e.g.,][]{2112.10990,10.3847/1538-4357/ac6acf,2501.01495},
where the signal amplitude is assumed as constant over the entire observation period.
The implementation in the \cwinpy package~\citep{10.21105/joss.04568} also allows for transient time-dependent profiles
as per Equations \eqref{eq:cw_windwo_rect} or \eqref{eq:cw_windwo_exp},
though all signals considered will have
fixed start time
(chosen here as $t_0=\Tgl$)
and only vary in $\tau$
(from 1\,hr to 120\,d).

To make the inference computationally tractable, the data are massively compressed \citep{10.1103/PhysRevD.72.102002}:
First, they are heterodyned to account for the motion of the detectors with respect to the source as well as any intrinsic phase evolution,
fixing to the best-fit parameters from Table~\ref{tab:Vela_glitch}.
These heterodyned data are then lowpass-filtered and down-sampled.
\cwinpy then uses the \texttt{bilby} \citep{2019ApJS..241...27A} and \texttt{dynesty} \citep{10.1093/mnras/staa278} packages for nested sampling inference to produce a Bayesian evidence value and posterior probability distribution for the unknown source parameters, given the compressed data.

Eight \cwinpy analyses were performed, toggling three different options:
(i) a rectangular or exponential amplitude profile;
(ii) unconstrained $\iota$ and $\psi$
or observationally motivated constraints on these angles
from modeling of the pulsar wind nebula \citep{2008ApJ...673..411N};
(iii) emission only from the $l=m=2$ mass quadrupole modes near $2\frot$, or allowing for both the $l=2$, $m=1,2$ modes, with emission at both the $\frot$ and $2\frot$ \citep{2010MNRAS.402.2503J, 2019ApJ...879...10A} harmonics.
See Appendix~\ref{app:cwinpy} for a full description of these options,
other prior choices,
and results.

No evidence for coherent signals across H1 and L1 was found.
The posterior probability distributions on signal amplitudes have been used to define 95\% credible upper limits as a function of the binned signal duration.

\subsection{Transient \texorpdfstring{\Fstat}{F-statistic} search}
\label{sec:tcw-fstat}

This method \citep{1104.1704}
builds on the standard matched-filter \cw maximum-likelihood \Fstat \citep{gr-qc/9804014}
in its \lalsuite \citep{lalsuite} frequency-domain implementation.
It has already been used in O2 and O3 post-glitch \gw searches \citep{1907.04717,2112.10990,2201.08785,2303.16720}.

We use 1800\,s \sfts \citep{T040164}
produced from time-domain gated data \citep{gating},
combined coherently
across both LIGO detectors \citep{10.1103/PhysRevD.72.063006} and the full observing time.
The detection statistic is
a Bayes factor $\BtSG$ marginalized over start times and durations \citep{1104.1704},
under the two amplitude profiles from
Equations \eqref{eq:cw_windwo_rect} and \eqref{eq:cw_windwo_exp},
and evaluated over a template bank
in $\frot$, $\fdotrot$, $\fddotrot$, and $\fdddotrot$.
Unlike \cwinpy,
this search does not use inclination and polarization constraints.
(They can be incorporated into an \Fstat framework as per \citet{2010CQGra..27s4015J} but this has not yet been implemented for transients.)

We allow for signal start times $t_0$ within $\pm1$\,day of $\Tgl$
and duration parameters $\tau$ up to 120\,days,
covering data from
28 April 2024
until
28 August 2024.
The search ranges are given in Table~\ref{tab:Fstat_setup} of Appendix~\ref{app:Fstat}.
Rectangular-window (constant amplitude) signals
can be searched for efficiently on CPUs with code in \lalsuite~\citep{lalsuite},
while exponentially decaying signals require the GPU implementation
in \pyfstat~\citep{1805.05652,2101.10915}.
Both are called via the \tcwip
(``transient CW investigation pipeline'')
package~\citep{tcwip}.

We find no statistically significant candidates
above a threshold set to the expected loudest outlier from
an empirical background estimation~\citep{2111.12032}.
We obtain upper limits on the initial strain amplitude $h_0$ via simulated signal injection and recovery.
See Appendix~\ref{app:Fstat} for details on the method, upper limits procedure and full results.

\subsection{Weighted peakmap search}
\label{sec:tcw-freqhough}

As a complementary approach to the two coherent searches, we also run a semi-coherent search based on \wpm.
This uses time-domain \bsd files \citep{Piccinni:2018akm}, including a glitch cleaning step~\citep{Astone:2005fj}.
The search starts
2000 seconds before the glitch time and covers $\Ttot=101.7$\,days.
Given the significant \asd difference of the L1 and H1 detectors, only L1 data are used.
Data over a 0.2\,Hz frequency band around the putative signal frequency are extracted from the \bsds
and heterodyned \citep{10.3847/1538-4357/ac6acf} to remove the Doppler modulation induced by the Earth's motion and the Vela pulsar's secular spin-down, using values measured immediately after the glitch, see Table~\ref{tab:FH_setup}.
The corrected data are divided in segments of duration $\Tseg$
(see Appendix~\ref{app:FHwin}),
and from each an equalized spectrum is computed.
Local maxima above a given threshold are selected,
creating a so-called \textit{peakmap} \citep{Astone:2014esa}.

We then search for peaks in the projection of the peakmap on the frequency axis,
as expected for sufficiently strong signals.
Peaks are weighted by their amplitude before projection, increasing the contribution of peaks from the early part of the signal, which is expected to be stronger.
The detection statistic
is the critical ratio 
\mbox{$CR=(\mathcal{P}-\mu_\mathcal{P})/\sigma_\mathcal{P}$},
where $\mathcal{P}$ is the sum of the weighted peaks
and $\mu_\mathcal{P}$, $\sigma_\mathcal{P}$ are the corresponding projection mean value and standard deviation.

This procedure is independent of the assumed signal model.
But by assuming a specific signal model we can make an ``optimal'' choice of the observation window $\Tobs\le \Ttot$ and segment duration $\Tseg$ --
see Appendix~\ref{app:FHwin}
and the resulting choices
in Table~\ref{tab:FH_summary}  for the seven considered search setups.

For each of these setups,
we select the ten most significant outliers as those with the highest $CR$ from each of ten 0.02\,Hz sub-bands.
Those with false-alarm probability
(taking into account the trials factor)
larger than 0.01 are discarded.
Three outliers remain:
one corresponding to $\tau=8$ days with frequency 22.455\,Hz ($CR=6.65$),
one to $\tau=30$ days with frequency 22.5011\,Hz ($CR=7.55$)
and one to $\tau=1000$ days with the same frequency ($CR=5.15$).
They are subjected to a follow-up procedure based on different methods, see Appendix~\ref{app:wpm_fu}.
While the last two are associated to an instrumental spectral line, the first one is more difficult to interpret, see Appendix~\ref{app:wpm_fu}. 
As it is statistically marginal, we do not consider it a promising detection candidate, and also proceed for this method to compute upper limits.

The upper limits are placed in terms of the initial signal strain amplitude
via simulated signal injections.
For the polarization parameters $\psi$ and $\iota$ we use both the restricted prior described for \cwinpy in Appendix~\ref{app:cwinpy} and an unrestricted uniform distribution.
See Appendix~\ref{app:FHwin} for details,
with Figure~\ref{fig:cw_uls_wpm} and Table~\ref{tab:FH_summary}
including the full results.

\subsection{Hidden Markov model search}
\label{sec:tcw-hmm}

The fourth search for transient quasi-monochromatic \gws combines the semi-coherent \Fstat \citep{gr-qc/9804014,lalsuite}
with \hmm tracking \citep{1606.02412, 1710.00460}
at both once and twice the Vela spin frequency \citep{PhysRevD.99.123010}.
It uses the same $1800\,\mathrm{s}$ \sfts as the transient \Fstat method.

Instead of a fully coherent analysis,
the \hmm pipeline divides the data into segments of duration $\Tcoh$ and evaluates the standard \Fstat over a template grid coherently in each segment
(without explicitly applying time-windowed search templates),
which are then recombined incoherently.

The \hmm signal model allows for a drift in frequency between each segment, i.e., for stochastic spin-wandering or other deviations from the standard Taylor phase model of Equation~\eqref{eq:timing-model}.
We employ the Viterbi algorithm \citep{1054010} to determine the optimal frequency path for both the single harmonic ($2\frot$) and dual harmonic ($\frot$ and $2\frot$) scenarios.
The detection statistic is the log-likelihood of the optimal path normalized by the number of segments,
$\bar{\mathcal{L}}\equiv\mathcal{L}/N_T$.

The configuration of the \hmm search is described in Table~\ref{tab:HMM_setup} in Appendix~\ref{app:hmm}.
It covers a $0.05\,\mathrm{Hz}$ band centered on $\fgw=\frot$ and $0.1\,\mathrm{Hz}$ band centered on $\fgw=2\frot$,
with frequency resolutions of $1/(4\Tcoh)$ and $1/(2\Tcoh)$, 
respectively.
The observation window covers $\Tobs=90$ days.
For added flexibility, we repeat the \hmm analysis with three values of $\Tcoh$, each corresponding to one of the three glitch recovery timescales.
Shorter $\Tcoh$ reduces sensitivity, but can track a more rapidly fluctuating signal.

Empirical detection thresholds are set using two methods
(see Appendix~\ref{app:hmm}),
based on either repeating the search on numerous realizations of Gaussian noise data, or repeating the search at randomly chosen off-target sky positions.

The full results of the \hmm search are compiled in Table~\ref{tab:HMM_results}, where we list the max-$\bar{\mathcal{L}}$ statistics obtained with single or dual harmonic tracking for each value of $\Tcoh$.
Our analysis does not recover any outliers with false-alarm probability below $1\%$.
We therefore find no statistically significant outliers for the \hmm search.

We set upper limits by injecting simulated transient signals with varying durations $\tau$ into the detector data, and then recovering the injection using the same \hmm search configurations,
with single or dual harmonic tracking. 
Upper limit results for the single harmonic emission ($2\frot$ only) scenario with $\Tcoh=9$ hr are included in Figure~\ref{fig:cw_uls},
with the full set of upper limits shown in Appendix~\ref{app:hmm}
and Figure~\ref{fig:h0_uls_hmm}.

\subsection{Comparison of upper limits results}
\label{sec:tcw-uls}

\begin{figure}[ht!]
\includegraphics[width=\columnwidth]{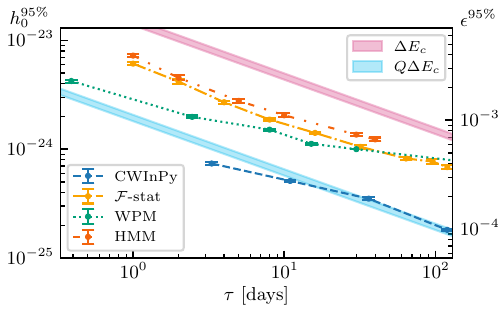}
\caption{ 
\label{fig:cw_uls}
Selected upper limits at 95\% confidence from the four \cw-like post-glitch searches,
in terms of initial strain amplitude (left-hand axis)
and \ns ellipticity $\epsilon$ (right-hand axis),
both defined at the start of signals
with exponentially decay parameters $\tau$.  \\
The results were chosen to illustrate how upper limits
scale with the strictness of prior assumptions built into each search configuration.
The \cwinpy results have the strictest constraints
(fixed frequency evolution and
constrained orientation angles),
while the three other results are for narrow-band searches,
with
\Fstat results marginalized over all durations,
\wpm results for search configurations optimized at each $\tau$ point,
and \hmm results for $\Tcoh=9$\,h. \\
For comparison, \gw emission at the characteristic energy scale $\Delta\Echar$ from Equation~\eqref{eq:E_characteristic} is shown as the magenta shaded band near the top,
while the sky blue shaded band further down shows the emission if the \gw energy output is suppressed by
a factor $Q\approx\Qvalue$, as in the \citet{2007.05893} transient mountain model.
}
\end{figure}

In Figure~\ref{fig:cw_uls} we compare a selection of upper limits from each of the four methods,
focusing on signals at twice the pulsar's rotation frequency and with exponential amplitude decay.
The results were chosen to span the range of assumptions of varying strictness
covered by the set of pipelines:
the strictest upper limits for such signals are obtained
by the fully-coherent targeted \cwinpy search with
constrained orientation angles,
while the semi-coherent \hmm search with the least prior assumptions provides the least strict constraints.
The two other methods and other configuration choices provide intermediate strictness of both assumptions and resulting constraints.
The figure also contains comparisons with
theoretical expectations on possible post-glitch \gw strains,
which will be discussed in the following section.

These limits concern arbitrary exponentially decaying \cw-like transients within the search ranges,
and at least for the \Fstat method
it is known that the exact amplitude evolution only has a small effect on its sensitivity \citep{1104.1704,1907.04717}.
In addition, from the \wpm method we also report a dedicated upper limit
for \gw amplitudes following the three-component frequency decay observed in the radio,
for which we find $h^{95\%}_0=4.5\times 10^{-24}$ assuming uniform priors on the polarization parameters and $4.7\times 10^{-24}$ for restricted priors.
(See also Table~\ref{tab:FH_summary} in the appendix.)

How deep a search has probed below the detector noise floor can be stated as a sensitivity depth~\citep{Behnke:2014tma,Dreissigacker:2018afk,Wette:2023dom}
\mbox{$\mathcal{D}=\sqrt{S_n(f)}/h_0$},
where $\sqrt{S_n}$ is the \asd suitably averaged over the observation time.
These searches have probed below the average four-month H1+L1 \asd by $\mathcal{D}$ factors
from 4\,Hz$^{-1/2}$ for the most flexible \hmm search (at $\tau=1$\,d)
to 170\,Hz$^{-1/2}$ for the most restrictive \cwinpy analysis (at $\tau=120$\,d).

\section{Interpreting the observational constraints}
\label{sec:interpretation}

We now provide some interpretation of our results.
In Section~\ref{sec:interpret_bursts} we discuss what conclusions can be drawn given the lack of detection of a burst-like signal. 
In Section~\ref{sec:interpret_transient} we discuss the same for the non-detection of long-duration transient quasi-monochromatic signals.
In Section~\ref{sec:interpret_joint} we discuss the combined constraints from both on \gw emission models and \ns properties.

Before looking at the separate burst and long-transient cases, it is useful to assemble some formulae relevant to both.
Despite searching for signals over very different frequencies and durations, both types of search cover signals of fixed source frequency and of finite duration.
Hence, we can also describe the burst analyses from Section~\ref{sec:burst}
(for the case of f-mode or similar signals)
with the same notation as for \cw-like signals from Section~\ref{sec:tcw},
using the explicit $h_0(t)$ and converting to the average $\hrss$ where needed.
The total \gw energy in such a signal is then given by
\citep[see e.g.\ equation (3) of][]{1104.1704}:
\begin{equation}
\label{eq:E_gw_general}
    \Egw = \frac{2\pi^2}{5} \frac{c^3}{G} \fgw^2 d^2 
    \int_{t_0}^{t_0 + \Tobs} h_0^2(t) dt \,,
\end{equation}
where $\Tobs$ is the observation time.
This applies regardless of the time variation of the signal amplitude $h_0(t)$.
The corresponding optimal signal-to-noise ratio is
\begin{equation}
\label{eq:rho_squared_general}
    \rho_0^2 = 
    \int_{t_0}^{t_0 + \Tobs} \frac{B(t;\hat{n},\iota,\psi) \, h_0^2(t)}{S_h(t,\fgw)} dt \,,
\end{equation}
where $S_h(t,\fgw)$ is the power spectral density of the detector noise
and $B(t;\hat{n},\iota,\psi)$ summarizes location- and orientation-dependent factors
\citep[for details see][]{gr-qc/9804014,1104.1704,T0900149}.

For a simple exponential decay on timescale $\tau$, the amplitudes $\hrss$ and $h_0$ are related by
\begin{equation}
\label{eq:h_rss-h_0}
    \hrss = h_0 \sqrt{\frac{\tau}{2}} ,
\end{equation}
as can easily verified by assuming an exponential time dependence for the integrand in Equation~\eqref{eq:h_rss_definition}.

We will now use these formulae to interpret the results from Sections \ref{sec:burst} and \ref{sec:tcw},
and compare with the characteristic energy scale of Equation~\eqref{eq:E_characteristic}
and additional model-specific constraints.

\subsection{Interpretation: burst results}
\label{sec:interpret_bursts}

The results of our burst searches were presented in Figure~\ref{fig:bursts-sensitivity}.
For the decaying exponentials assumed in the injections to set these limits, Equation~\eqref{eq:rho_squared_general} reduces to
\mbox{$\rho_0^2 \propto \hrss^2 / S_h(\fgw)$},
defining $S_h(\fgw)$ as a time average over $S_h(t,\fgw)$.
Hence, at fixed signal-to-noise ratio, the minimum detectable $\hrss$ scales as $\propto\sqrt{S_h(\fgw)}$\,.
This accounts for the limits weakening with increasing frequency,
reflecting the slope of the detector noise curve as seen in Figure~\ref{fig:gw_data}

To interpret the astrophysical significance of the results, we can similarly re-write Equation~\eqref{eq:E_gw_general} to give
\begin{equation}
\label{eq:E_gw_h_rss}
    \Egw = \frac{2\pi^2}{5} \frac{c^3}{G} \fgw^2 d^2 \hrss^2 \,.
\end{equation}
Setting $\Egw$ equal to the characteristic energy of Equation~\eqref{eq:E_characteristic} then gives the corresponding $\hrss$ as a function of $\fgw$, with scaling \mbox{$\hrss \propto 1 / \fgw$}.
Inserting values for the \glitchdate Vela glitch,
we obtain 
\mbox{$\hrss(\fgw) \approx 9.7 \times 10^{-20} \, \mathrm{Hz}^{-1/2} / (\fgw/\mathrm{Hz})$},
which is plotted as the dashed green line in Figure~\ref{fig:bursts-sensitivity}.

The search upper limits constrain the actual \gw energy to be less than this at sufficiently low frequencies.
For instance, for the \xpipe this characteristic energy scale is surpassed for frequencies below about 800\,Hz,
and at the lowest injection frequency of 290\,Hz,
$\hrss$ is constrained to a factor of $\approx3$ below the characteristic scale.
Given the scaling \mbox{$\Egw \sim \hrss^2$} of Equation~\eqref{eq:E_gw_h_rss}, this corresponds to a factor of $\approx 9$ in radiated energy.
However, emission at such low frequencies would not correspond to the f-modes considered most plausible to be excited by pulsar glitches.
For example, see \citet{2403.09489}
where f-mode emission was studied for a catalog of 15 equations of state
(eight for standard \nss and seven for quark stars).

Our upper limits on $\hrss$ can be converted into upper limits on a (dimensionless) amplitude $\alpha$ of the f-mode excitation, such that a point on the surface of the \ns oscillates about its equilibrium position by a distance $\sim \alpha R$, with $R$ the stellar radius.
This conversion depends on the \ns mass $M$ and equation of state.  To provide a quantitative example of this, we will make use of the analytic results given in \citet{2204.12869} for uniform density stars. Combining equations (5) and (10) of \citet{2204.12869}, and making use of our Equation~\eqref{eq:h_rss-h_0}, one obtains
\begin{equation}
    \alpha = \left(\frac{2\pi}{15}\right)^{1/2}   \left(\frac{c^3}{G}\right)^{1/2}
    \frac{\hrss \, d}{M^{1/2} R} \,.
\end{equation}
Inserting $M= 1.4\,M_\odot$ and $R = 10^6$\,cm, and the distance for Vela, we find
\begin{equation}
    \alpha \approx 7 \times 10^{-5} \, 
    \left(\frac{\hrss}{3 \times 10^{-22} \, {\rm Hz}^{-1/2} } \right) ,
\end{equation}
where we have chosen the scaling in $h_{\rm rss}$ based on the upper limits on $\hrss$ of Figure~\ref{fig:bursts-sensitivity},  at a frequency of about $1.5$\,kHz, typical for an f-mode.
Roughly speaking, we have constrained the oscillation amplitude $\alpha R$ of a particle on the Vela's surface to be no larger than about $70$\,cm for a typical f-mode frequency.
However, at these frequencies our upper limits on $\hrss$ fall short of the values corresponding to the characteristic energy by about a factor of $\sim 4$.

\subsection{Interpretation: long transient results}
\label{sec:interpret_transient}

As discussed in Section~\ref{sec:tcw-emission}, there are two main mechanisms for producing \cw-like transients following a glitch.
The first we discussed was Ekman pumping.
Using equation (55) of \citet{1008.0236} and the parameters for the \glitchdate Vela glitch, we obtain an estimate of the signal strength of
\mbox{$h_0 \approx 7 \times 10^{-26}$}.
While this estimate is conservative, mainly due to uncertainties on the lag between the components in the \ns interior
(which may be larger than the observed glitch size, \citealt{1502.07062}),
and unknown nuclear physics parameters that enter the model, the predicted strain is significantly lower than the upper limits on $h_0$ from our analyses
(Figure~\ref{fig:cw_uls}),
which are hence not yet sensitive enough to set constraints on this mechanism.

The second emission mechanism we discussed was the formation of a transient mountain that then relaxes away \citep{2007.05893}.
To help interpret results from Figure~\ref{fig:cw_uls} under this mechanism, we can insert the assumption of exponential decay on a timescale $\tau$, Equation~\eqref{eq:cw_windwo_exp}, as would be expected for dissipative mountain decay,
into Equation~\eqref{eq:E_gw_general} for $\Egw$.
If we assume \mbox{$\tau \ll \Tobs$}, we obtain 
\begin{equation}
\label{eq:E_gw_exponential}
    \Egw = \frac{\pi^2}{5} \frac{c^3}{G} \fgw^2 d^2 h_0^2(0) \tau \,.
\end{equation}
For a fixed $\Egw$, this shows that \mbox{$h_0 \sim \tau^{-1/2}$}, explaining the slope of the upper limits in the figure.  

To understand the astrophysical significance of these results,
we can use Equation~\eqref{eq:E_tCW_with_Q}
where only a fraction $Q$ of the characteristic glitch energy of Equation~\eqref{eq:E_characteristic} goes into \gw emission.
This yields for the initial \gw strain amplitude
\begin{equation}
 \label{eq:tcw_indirect_UL_PGM}
 h_0^{Q\Delta\Echar} \approx \frac{1}{d} \sqrt{\frac{5G}{2c^3} \frac{I}{\tau} Q \frac{\Delta \frot}{f}} \,.
\end{equation}
For $Q=1$, this corresponds to
what was discussed as ``indirect energy limits'' in \citet{1104.1704,1907.04717,2112.10990},
and is shown (as a function of $\tau$) as the magenta band in Figure~\ref{fig:cw_uls}.
The
width of the band includes uncertainties on the Vela pulsar distance and its timing parameters,
but not on the \ns moment of inertia which could vary by up to another factor of 3 for conventional models
or 5 for extreme matter compositions~\citep{Lattimer:2021emm,Johnson-McDaniel:2012wbj}.
The observational \gw upper limits from all analyses are stricter than this by factors of $\sim 3$ to $\sim 10$.
Notwithstanding the phenomenological nature of this indirect benchmark, this is a significant milestone for post-glitch transient \gw analyses.

To interpret the results in terms of a more specific model, we consider the \citet{2007.05893} prediction of the energy fraction $Q$ corresponding to the healing parameter defined in Equation~\eqref{eq:healingQ}.
The \glitchdate Vela glitch had three observed recovery components
(see Equation~\ref{eq:glitch-model}),
but to be
consistent with the other quantities in Figure~\ref{fig:cw_uls},
we show as the sky-blue band
a simplified result,
approximating the logic of the model:
the result of Equation~\eqref{eq:tcw_indirect_UL_PGM}
with \mbox{$Q\approx\Qvalue$} and as a function of a single $\tau$.

Most upper limits are still above this prediction, while the result from the most constrained \cwinpy search just reaches this level of emission.
So under optimal assumptions,
a detection would have been just about possible under this model,
depending on the assumed \gw signal duration $\tau$
and the actual moment of inertia $I$ of the Vela \ns.
The quantitative implications of non-detection under this model are considered in Section \ref{sec:interpret_joint}.

We can also recast our results in terms of the additional ellipticity $\epsilon$ imparted to the Vela pulsar right after the glitch, which is related to the \gw amplitude by the equation \citep[see e.g.][]{2206.06447}
\begin{equation}
    h_0 = \frac{4 \pi^2 G \epsilon I \fgw^2}{c^4 d} \,.
\end{equation}
When applied to Vela this gives
\begin{equation}
    \epsilon \approx 5.4 \times 10^{-4} \left(\frac{h_0}{10^{-24}}\right) \,,
\end{equation}
interpreted as the initial transient ellipticity increase that then decays away again.

From Figure~\ref{fig:cw_uls}, we see that our limits on $h_0$ span the range
$[2\times10^{-25},8\times10^{-24}]$,
depending on the search and on $\tau$,
corresponding to an ellipticity range
of $[10^{-4},3\times10^{-3}]$.
In an astrophysical context, these are very large values---about three orders of magnitude larger than the largest mountains that a \ns crust is estimated to be able to support \citep[see e.g.][]{Johnson-McDaniel:2012wbj,2018ASSL..457..673G}.
However, such estimates assume a \emph{persistent} (i.e.\ infinitely long-lived) mountain.
A realistic crust may respond \emph{plastically} to the stresses that support such a mountain \citep{2018MNRAS.480.5511B}, so that the exponential post-glitch relaxation may represent the plastic decay of a larger asymmetry.
For this reason, the upper limits we have obtained are of astrophysical interest, despite being larger than the theoretical ones for permanent mountains.
For further comparison, the \cwinpy upper limit on a persistent ellipticity of the Vela pulsar from 8 months of O4a data
was $\epsilon \leq 4.7\times10^{-5}$, also at 95\% confidence \citep{2501.01495}.

As discussed in Section~\ref{sec:tcw-emission},
if the glitch caused the Vela pulsar to go into transient free precession,
there would be additional emission at \mbox{$\fgw\approx\frot$}.
Due to the steep detector noise curve at low frequencies,
strain upper limits are less constraining by an order of magnitude in this regime.
Quantitative details are given in appendices \ref{app:cwinpy} and \ref{app:hmm}.

\subsection{Interpretation: joint analysis and NS properties}
\label{sec:interpret_joint}

We now show how
our non-detections can be used to make statements
about the \ns parameter ranges under which specific \gw emission models can be valid.
Since this is the first time that both short- and long-duration \gw searches have been performed for the same pulsar glitch,
we aim to illustrate the benefit of considering results from the two types of searches jointly.
So we assume that both a burst signal and a longer-duration quasi-monochromatic transient were produced at the glitch.
The fact that these signals were \emph{not} detected either
(i) rules out emission models that predict \gw energy fractions in one of the signal channels that our searches would have detected,
or (ii) places non-trivial constraints on the \ns mass and radius
in the sense that Vela could still have emitted \gw energy as predicted by the models,
but at parameters leading to reduced detectability.

Following the method presented by \citet{ball_joint_implications},
we assume specific models for both short- and long-duration waveforms. 
For bursts,
we assume a rapidly damping sinusoid,
consistent with an f-mode,
in the form used by \citet{2003.12082} and parameterized by $\hrss$,
as discussed in Section~\ref{sec:burst-emission}.
Meanwhile,
the long-transient model corresponds to the decay of a transient mountain \citep{2007.05893},
as discussed in Section~\ref{sec:tcw-emission},
with an initial strength $h_0$.

\begin{figure}
    \centering
    \includegraphics[width=\columnwidth]{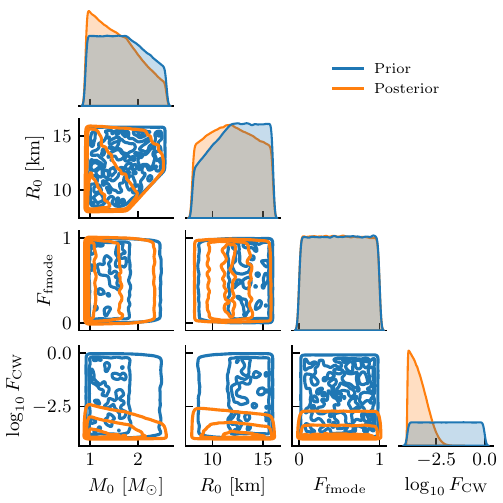}
    \caption{Prior and posterior distributions of \ns mass, radius
    and energy fractions $\digamma_{\mathrm{fmode}}$ and $\log_{10}(\digamma_{\mathrm{CW}})$
    under the assumption that \gws from both a short-duration f-mode and a long-duration transient mountain scenario were emitted but not detected.
    After including the search results, there is little effect on the allowed f-mode energy in the physically motivated region $\digamma_{\mathrm{fmode}} < 1$ but a measurable effect on the allowed \cw energy ($\digamma_{\mathrm{CW}}$).
    Contours show $10\%$, $50\%$, and $90\%$ credible intervals.
    \label{fig:MR_corner_plot_variable_fmode_CW}
    }
\end{figure}

\begin{figure}
    \centering
    \includegraphics[width=\columnwidth]{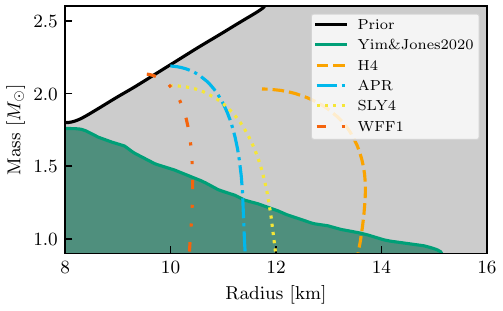}
    \caption{\ns mass and radius regions
    for which the Vela pulsar could have emitted \gws according to the \citet{2007.05893} model after its 2024 glitch, but without our searches detecting them (within the green shaded region at the lower left, at $90\%$ confidence).
    This is compared to the assumed prior ranges (larger black contour)
    and to mass--radius curves for some example equations of state \citep{wiringa_equation_1988,PhysRevC.58.1804,douchin_unified_2001,PhysRevD.73.024021},
    for which allowed \ns configurations would lie along each curve.
    Here we fixed \mbox{$\digamma_{\mathrm{CW}}=\Qvalue$} to match the healing parameter $Q$ defined in Equation~\eqref{eq:healingQ} 
    and arbitrarily chose \mbox{$\digamma_{\mathrm{fmode}}=0.8$}
    (which has little effect).
    We see that the Yim \& Jones model is only compatible with our non-detection for smaller mass and radius ranges.
    \label{fig:mass_vs_radius_fixed_fmode_variable_CW}
    }
\end{figure}

We parameterize $\Egw$ in each channel as some fraction $\digamma$ of the energy $\Delta\Echar$
from Equation~\eqref{eq:E_characteristic}.
This gives
\mbox{$\Egw = E_{\mathrm{CW}}+E_{\mathrm{fmode}} = \digamma_{\mathrm{CW}} \Delta\Echar + \digamma_{\mathrm{fmode}} \Delta\Echar$}.
We employ an equation-of-state independent universal relation from \citet{1608.02582} to connect the pulsar's moment of inertia $I$
to the non-rotating mass and radius.
We use another set of equation-of-state independent universal relations from \citet{PhysRevC.106.015805} to connect the f-mode frequency and damping time to the non-rotating mass and radius. 

We perform Monte Carlo simulations of the parameter space, drawing samples from a set of prior distributions.
For each prior sample, the initial CW-like signal amplitude $h_0$ and the f-mode integrated strain $\hrss$ are calculated.
These values are compared to detection efficiency curves of a search pipeline for simulated waveforms,
i.e., the probability of the search recovering a signal with some morphology at different amplitudes.
The priors
are described in Appendix~\ref{app:joint},
along with how the likelihood function
is constructed from the \xpipe analysis and the most constrained \cwinpy configuration
as representatives of the two search types.

We first let $\digamma_{\mathrm{fmode}}$ and $\digamma_{\mathrm{CW}}$ independently vary up to a maximum of $1$.
The posterior distribution for mass, radius, $\digamma_{\mathrm{fmode}}$, and $\digamma_{\mathrm{CW}}$ is shown in Figure~\ref{fig:MR_corner_plot_variable_fmode_CW}.
Consistent with the discussion above of the relative strictness of the upper limits from both types of searches in comparison to the $\Delta\Echar$ energy scale, burst search limits have little to no effect on the posterior distribution,
while \cw search limits strongly constrain $\digamma_{\mathrm{CW}}$.
From this posterior,
we constrain $\digamma_{\mathrm{CW}}$ to below $5.65\times10^{-3}$ with $95\%$ confidence.

Given the minimal effect of $\digamma_{\mathrm{fmode}}$, we then fix it to an arbitrary value of $0.8$ and explore
the allowed mass and radius values
if we assume the \citet{2007.05893} model correctly predicts $\digamma_{\mathrm{CW}}$, yet no detection was made.
Figure~\ref{fig:mass_vs_radius_fixed_fmode_variable_CW}
shows how setting $\digamma_{\mathrm{CW}}$ equal to the healing parameter {$Q\approx\Qvalue$} defined in Equation~\eqref{eq:healingQ} requires a smaller mass and radius for the Vela pulsar if \gw emission as, per this model, would have been present after this glitch.
This effect is driven by the moment of inertia relation from \citet{PhysRevC.106.015805}.

In addition to being dependent on the specific emission model considered and the tightest prior assumptions among the set considered in our analyses, these results are still less informative compared to other \ns mass--radius limits from the binary \ns merger GW170817 \citep{abbott_gw170817_2017,abbott_gw170817_2018},
observations of \nss in binary systems \citep[e.g.,][]{taylor_new_1982,barr_pulsar_2024},
or NICER pulse profile modeling \citep[e.g.,][]{riley_nicer_2021,raaijmakers_constraints_2021}.

Applying this method to future glitches with improved \gw detector sensitivity,
a continued lack of detection could place stronger constraints on models such as \citet{2007.05893},
allow us to probe different emission scenarios,
and to gain stronger benefits from a joint burst--CW analysis.

\section{Conclusions}
\label{sec:conclusions}

Pulsar glitches are unique probes of the structure and dynamics of \nss.
The improved sensitivity of the LIGO detectors during the O4 run
and the detailed timing information from the IAR and MPRO radio telescopes
have now allowed us,
for the first time,
to place physically meaningful
upper limits
on the \gw emission from a glitching pulsar,
constraining it to below the indirectly inferred energy scale
for at least part of the relevant parameter space.

Informed by an ephemeris fit to the combined IAR--MPRO data,
we have analyzed LIGO data around and after the \glitchdate glitch
of the Vela pulsar,
one of the closest and most prolific glitching pulsars known.
None of the \gw searches found promising detection candidates.

Our search with three unmodelled methods for \gw bursts
covering different signal durations from milliseconds to minutes
has set constraints on the \gw strain
that up to 700\,Hz are,
for the first time,
stricter than the values corresponding to the full glitch excess energy.
The tightest constraint is 
$\hrss \approx 8.6 \times 10^{-23}$
at 290\,Hz,
a factor of $\sim3$ below the characteristic energy scale.

We also searched for quasi-monochromatic long-duration transients
of up to four months
near twice the Vela pulsar's rotation frequency.
With four methods of varying strictness in their assumptions,
we have excluded \gw strain as low as a factor of 10
below the characteristic scale
if all energy liberated at the glitch
were converted into exponentially decaying quasi-monochromatic \gws
and under the strictest assumptions,
and still put constrains below that energy scale even under relaxed assumptions and phase evolution models.
We also searched for similar signals from free precession at once the rotation frequency,
but results are less constraining due to the steep rise of the LIGO noise curve towards lower frequencies.

Previous searches for \gw bursts from the Vela pulsar \citep{10.1103/PhysRevD.83.042001},
all-sky \gw burst searches \citep{2507.12282,2507.12374},
and quasi-monochromatic \gw transient searches for Vela and other glitching pulsars \citep{1907.04717,2112.10990,2201.08785,2303.16720}
all still yielded results above the characteristic glitch energy scale.
In contrast, our new results for the first time allow meaningful
comparisons with specific emission models.

Under the burst scenario,
we can constrain emission only at lower frequencies but not yet
at a kHz or above, which most \ns equations of state predict~\citep[see, e.g.,][]{gr-qc/9711088,2403.09489},
but where detector sensitivity is worse,
primarily being limited by laser shot noise \citep{2411.14607}.

Under the ``transient mountain'' scenario of \citet{2007.05893},
the scale of long-duration emission
is set by the transient part of the glitch's spin-down change,
which in the case of this Vela glitch
is unfortunately a small fraction,
and hence we only reach the sensitivity to probe this model's predictions under the most constraining prior assumptions in one of our searches.

Combining the most constraining burst and long-duration search results, respectively,
we have also illustrated how a joint analysis can constrain the \ns mass and radius ranges consistent with specific emission models under a non-detection of post-glitch \gws.
However, further improvement particularly in the f-mode upper limits will be required for the full benefits of this method.

With the sensitivity of the \lvk detector network further improving
towards its fifth observing run
\citep{1304.0670}
and the Vela pulsar producing large glitches
every two to three years,
we can expect further improvements to this type of analysis
in the coming years.
Other glitching pulsars can also be promising targets, 
though the Vela pulsar stands out due to its low distance and strong glitches \citep{2206.14515,2210.09907,2406.00283}.
Going beyond the non-detection constraints we introduced here,
a positive detection of \gws from a glitching pulsar
would be another important milestone
in \ns astrophysics.

Data products associated with this paper are released at \url{https://doi.org/10.5281/zenodo.17735648}.

\section{Acknowledgments}
\input{P2000488_v28}
This article has been assigned document number \dcclink.

%

\vspace{5mm}
\facilities{LIGO, MtPO:26m, IAR}







\software{
\texttt{astropy} \citep{2013A&A...558A..33A,2018AJ....156..123A},
\cwb~\citep{Klimenko:2015ypf,Drago:2020kic,Martini:2025ewk},
\cwinpy~\citep{10.21105/joss.04568},
\texttt{distromax}~\citep{2111.12032,distromax},
\texttt{DSPSR}~\citep{10.1071/AS10021},
\texttt{gwpy}~\citep{gwpy},
\lalsuite~\citep{lalsuite},
\pyfstat~\citep{2101.10915,pyfstat},
\texttt{matplotlib}~\citep{matplotlib},
\texttt{numpy}~\citep{numpy},
\texttt{PRESTO}~\citep{2011ascl.soft07017R}, \texttt{PSRCHIVE}~\citep{10.1071/AS04022},
\pystampas~\citep{2021PhRvD.104j2005M},
\texttt{ROOT}~\citep{Brun:1997pa},
\texttt{scipy}~\citep{scipy},
\texttt{scikit-image}~\citep{scikit-image},
\texttt{SWIGLAL}~\citep{swiglal},
\tcwip~\citep{tcwip},
\texttt{TEMPO2} \citep{10.1111/j.1365-2966.2006.10302.x,astro-ph/0607664},
\xpipe~\citep{Sutton:2009gi,Was:2012zq}.
}

\appendix

This appendix includes additional details
on the setups and results of the four \cw-like long-duration post-glitch searches covered in Section~\ref{sec:tcw} of the main text,
as well as of the joint analysis of burst and \cw results covered in Section~\ref{sec:interpret_joint}.

\section{Details of the \texorpdfstring{\cwinpy}{CWInPy} analysis}
\label{app:cwinpy}

As mentioned in Section~\ref{sec:tcw-cwinpy},
eight search configurations were used for fully-coherent targeted time-domain Bayesian analyses with the \cwinpy package,
corresponding to all combinations of toggling three options. These options are described here in more detail, 
followed by the full results.

The first option allows the \gw strain amplitude to have either a rectangular or exponential decay profile.
The rectangular analysis uses a uniform prior on its duration  $\tau$
ranging from $\tau_{\rm min}=1$\,h to $\tau_{\rm max}=120$\,d, i.e.,
$p(\tau)
= (\tau_{\rm max} - \tau_{\rm min})^{-1}$.
The prior on the exponential profile's decay factor $\tau$ has the same range.
In both cases, $t_0$ is fixed at the glitch epoch. 

The second option either allows the orientation angle $\iota$ and polarization angle $\psi$ of the source to be unconstrained over their physically allowable ranges or to use observationally motivated constraints on their prior ranges derived from modeling of the pulsar wind nebula \citep{2008ApJ...673..411N}.
The angle $\iota$ corresponds to the inclination of the rotation axis with respect to the line of sight,
and $\psi$ measures the orientation of the rotation axis in the plane perpendicular to the line of sight.
For the unconstrained case, the prior distributions for both are uniform over the surface of a sphere,
such that $p(\iota) = \sin{\iota}$ over the range $[0, \pi)$ and $p(\psi) = 2 \pi^{-1}$ over the range $[0, \pi/2)$.
When constrained, the prior distribution for $\iota$ is a double-peaked Gaussian distribution with peaks at 1.1048 and 2.0368 radians and $\sigma=10.5$\,mrad,
and the prior distribution for $\psi$ is a Gaussian distribution centered at 2.2799 radians and $\sigma=1.5$ mrad\,\citep[see Appendix B of][]{2017ApJ...839...12A}.

The third option covers the cases of a search assuming emission only from the $l=m=2$ mass quadrupole mode at only twice the rotation frequency, or a search with potential emission from both the $l=2$, $m=1,2$ modes, with dual-harmonic emission at both once and twice the rotation frequency \citep{2010MNRAS.402.2503J, 2019ApJ...879...10A}.
In the latter case there is a characteristic amplitude for each mode, $C_{21}$ and $C_{22}$, both of which will have the same decay profile applied to them.
This also requires the input data to be processed via heterodyning at both frequencies.

\begin{figure}
    \centering
    \includegraphics[width=0.495\linewidth]{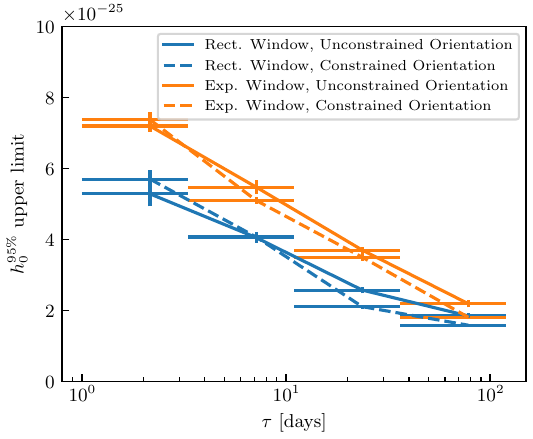}
    \caption{
    Upper limits from the fully-coherent targeted time-domain Bayesian search (\cwinpy) in terms of
    $h_0^{95\%}$
    as a function of signal duration $\tau$.
    This search assumes emission only from the $l = m = 2$ mass quadrupole mode.
    ``Rect. Window'' lines correspond to the time dependence given in Equation~\eqref{eq:cw_windwo_rect},
    and ``Exp. Window'' lines to that in Equation~\eqref{eq:cw_windwo_exp}.
    The solid and dashed lines indicate searches with unconstrained and constrained orientation angles, $\iota$ and $\psi$, respectively.
    \label{fig:cw_uls_cwinpy_h0}
    }

    \centering
    \includegraphics[width=\linewidth]{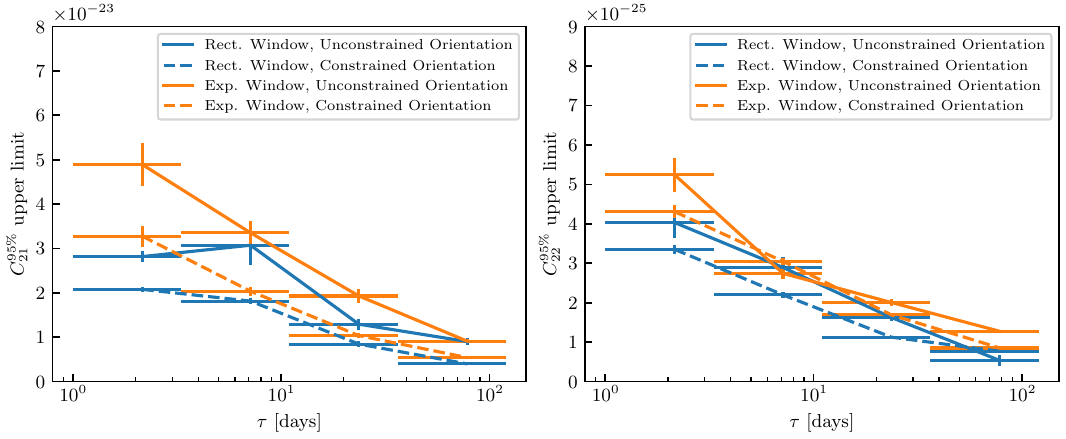}
    \caption{
    Upper limits from the time-domain Bayesian search (\cwinpy) in terms of $C_{21}^{95\%}$ and $C_{22}^{95\%}$ as a function of signal duration, $\tau$.
    This search assumes emission from the $l = 2, m = 1,2$ mass quadrupole modes, where the first
    panel
    contains searches for $C_{21}^{95\%}$
    and the second
    panel
    for $C_{22}^{95\%}$.
    ``Rect. Window'' lines correspond to the time dependence given in Equation~\eqref{eq:cw_windwo_rect},
    and ``Exp. Window'' lines to that in Equation~\eqref{eq:cw_windwo_exp}.
    The solid and dashed lines indicate searches with unconstrained and constrained orientation angles, $\iota$ and $\psi$, respectively.
    \label{fig:cw_uls_cwinpy_c21_c22}
    }
\end{figure}


In all cases, we use a broad unphysical uniform prior on the amplitude $h_0$, or on both $C_{21}$ and $C_{22}$ for the dual-harmonic mode \citep[see equations (1) and (2) of][]{10.3847/1538-4357/ac6acf}, bounded in the range $[0, 10^{-21})$.
The upper range is chosen to be large enough that the likelihood will be negligible at this point, while otherwise allowing the likelihood to dominate the posterior.
This allows to provide upper limits that are based on the data only.

Furthermore, in each search configuration,
data from the H1 and L1 detectors were analyzed
both coherently
and incoherently (i.e., with the H1 and L1 datasets combined or analyzed separately), 
providing Bayesian evidence values for a coherent signal between detectors and independent signals in each detector.
These evidences,
along with that from assuming purely Gaussian noise in both detectors,
have been combined to give an odds ratio
comparing a coherent multi-detector signal to incoherent signals in each detector or noise
\citep[see, e.g.][]{2017ApJ...839...12A}.

None of the results indicate more evidence for a coherent signal than for either an incoherent signal or none at all,
with the highest $\log_{10}$ Bayesian odds only reaching $-3.45$ from the configuration with exponential decay, unconstrained $(\iota,\psi)$, $l=m=2$ mode only. 
Full results from all configurations are included in the data release for this paper.

Due to the absence of evidence for coherent signals,
we
place upper limits from the posteriors on
\gw signal amplitudes
as a function of $\tau$ for each search configuration.
Figures \ref{fig:cw_uls_cwinpy_h0} and  \ref{fig:cw_uls_cwinpy_c21_c22} show upper limits
$h_0^{95\%}$ for the case where $l = m = 2$,
and $C_{21}^{95\%}$ and $C_{22}^{95\%}$ in the case for $l = 2, m = 1,2$,
respectively.
In each figure, results are included from analyses assuming the rectangular time dependence from Equation~\eqref{eq:cw_windwo_rect} or the exponential dependence from Equation~\eqref{eq:cw_windwo_exp},
and for unconstrained and constrained orientation angles, $\iota$ and $\psi$.
All these upper limits were calculated using samples from the posterior distributions constrained within the $\tau$ bins indicated by the horizontal bars in each figure.
The \cwinpy line in Figure~\ref{fig:cw_uls} corresponds to the exponential window and constrained orientation results,
i.e., the dashed orange line in the first panel of Figure~\ref{fig:cw_uls_cwinpy_h0}.

\section{Details of the transient \texorpdfstring{$\F$}{F}-statistic analysis}
\label{app:Fstat}

Here we provide additional details of the setup
and results of the transient $\F$-statistic search,
as summarized in Section~\ref{sec:tcw-fstat} of the main text.
For O4,
the search setup, distributed job management and post-processing
from the previous O2 and O3 searches \citep{1907.04717,2112.10990,2201.08785,2303.16720}
have been refactored as the \tcwip
(``transient continuous-wave investigation pipeline'')
package~\citep{tcwip}.
This wraps CPU code in \lalsuite~\citep{lalsuite}
for rectangular-window (constant amplitude) signals
and a \texttt{PyCUDA} GPU implementation for exponentially decaying signals
in \texttt{PyFstat}~\citep{1805.05652,2101.10915}.

The same grid of templates
is used to cover the frequency and frequency derivatives (up to $\fdddot$) ranges for both amplitude evolution options.
The grid spacings are calculated to yield a maximum metric mismatch \citep{gr-qc/0606088,0802.3332} of 0.02 over the full $\Tobs$.
The search ranges in all relevant parameters
summarized in Table~\ref{tab:Fstat_setup}.

To determine whether there are significant candidates from one of the two analyses,
we use the empirical background estimation procedure~\citep{2111.12032} implemented in the \texttt{distromax} package~\citep{distromax}:
We divide the search results into batches of 5000 templates,
fit the distribution of the per-batch maximum detection statistics,
and extrapolate this to obtain an estimate of the expected loudest outlier
from the full template bank.
The automated notching also described in \citet{2111.12032} is used
to reduce the impact of moderate-strength noise disturbances,
with one iteration step for the rectangular-window search results
and two steps for the exponential-window search results.
For rectangular signals,
the threshold is set at \mbox{$\log_{10}\BtSG^{\mathrm{thr}}\approx10.02$}
while the loudest candidate has \mbox{$\max\log_{10}\BtSG\approx9.65$}.
For the exponential-window search, the threshold is
set to \mbox{$\log_{10}\BtSG^{\mathrm{thr}}\approx9.30$}
and the loudest candidate has \mbox{$\max\log_{10}\BtSG\approx9.20$}.

\begin{deluxetable}{lr}
\tablecaption{
 \label{tab:Fstat_setup}
 Setup for the transient \Fstat search.
}
\tablehead{ \colhead{parameter} & \colhead{value}}
\startdata
$T_\mathrm{ref}$ [s] & 1398372935 \\
$t_{0,\min}$ [s] & 1398372935 \\
$\Delta t_0$ [d] & 2 \\
$\tau_{\min}$ [s] & 3600 \\
$\Delta \tau$ [d] & 120 \\
$\fgw$ [Hz] & 22.35453888321711 \\
$\Delta \fgw$ [Hz] & 0.022365721744089156 \\
$\fdot$ [Hz\,s$^{-1}$] & $-3.12874187125924\times10^{-11}$ \\
$\Delta\fdot$ [Hz\,s$^{-1}$] & $2.061675716980504\times10^{-13}$ \\
$\fddot$ [Hz\,s$^{-2}$] & $1.213100115859692\times10^{-21}$ \\
$\Delta\fddot$ [Hz\,s$^{-2}$] & $2.639553012031674\times10^{-23}$ \\
$\fdddot$ [Hz\,s$^{-3}$] & $2.5365827014929\times10^{-28}$ \\
$\Delta\fdddot$ [Hz\,s$^{-3}$] & $5.940719999999962\times10^{-30}$ \\
$\Ntemp$ & 757016 \\
\enddata
\tablecomments{
 The reference time $\Tref$
 for the \gw frequency and spin-down parameters $\fgw$, $\fdot$, etc.,
 is the first signal start time considered $t_{0,\min}$,
 which in turn is
 set to the earliest \sft data timestamp
 in $\Tgl\pm\min(\Delta\Tgl,1\,\mathrm{d})$.
 The minimum signal duration $\tau_{\min}$ is, for implementation reasons~\citep{T0900149},
 set to two \sft lengths.
 $\Delta t_0$ and $\Delta \tau$ are the ranges of start times and durations considered.
 The minimum frequency $\fgw$ and frequency bandwidth $\Delta \fgw$,
 and similarly for the three frequency derivatives,
 are set from the pulsar ephemeris
 (Table~\ref{tab:Vela_glitch}),
 using at each derivative order the maximum of
 (i) 0.001 times the central value of the parameter,
 (ii) $3\sigma$ uncertainties on it,
 (iii) for $\fgw$ and $\fdot$: the glitch step size.
 See \citet{2201.08785} for details on this setup procedure.
 $\Ntemp$ is the total number of frequency-evolution templates.
}
\end{deluxetable}

\begin{figure}[ht!]
\centering
\includegraphics[width=0.495\linewidth]{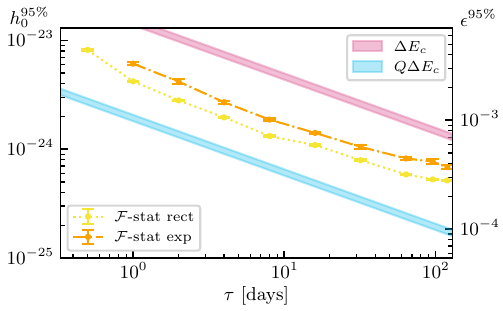}
\caption{
\label{fig:cw_uls_fstat}
Upper limits from the transient $\F$-statistic search
in terms of strain amplitude $h_0^{95\%}$ (left-hand axis)
and \ns ellipticity $\epsilon^{95\%}$ (right-hand axis),
as a function of the signal duration parameter $\tau$.
Results are included both
for signals with constant amplitude (``rect''),
as per Equation~\eqref{eq:cw_windwo_rect},
and for exponentially decaying signals (``exp''),
as per Equation~\eqref{eq:cw_windwo_exp}.
The same indirect energy limits are shown for comparison
as in Figure~\ref{fig:cw_uls},
corresponding to Equation~\eqref{eq:tcw_indirect_UL_PGM}
with $Q=1$ (magenta)
and $Q\approx\Qvalue$ (sky blue).
}
\end{figure}

We then obtain Frequentist 95\% upper limits on the initial strain amplitude $h_0$ by simulating signals
with rectangular windows or single exponential decay windows,
following Equations \eqref{eq:cw_windwo_rect} or \eqref{eq:cw_windwo_exp}.
That is, these upper limits are not making statements about a specific model following the three-component decay observed in the radio,
but about arbitrary \cw-like transients within the search range
as long as they fall within the range spanned by frequency and amplitude evolution templates.
In fact, the effects of different amplitude evolutions on the sensitivity are relatively minor, see \citep{1104.1704,1907.04717},
but still notable enough to motivate the use of two separate search configurations for rectangular and exponential window functions.

For each window, the simulated signals cover a set of discrete $\tau$ values to sample the $h_0$ upper limits curves, as shown in Figure~\ref{fig:cw_uls_fstat}.
At each $\tau$ value, a discrete set of $h_0$ values is covered to provide a detection efficiency curve, with 50 injections at each $h_0$ step.
For the injections,
the start time $t_0$ is randomly drawn from the $\pm1$ day search range,
the frequency evolution parameters are drawn uniformly from the same ranges as covered by the search,
and orientation angles are also uniformly sampled.
For each $h_0$,
we then count as detectable those injections that exceed the threshold
from the search with the corresponding window.

A sigmoid is fitted to the efficiency curve with \texttt{scipy}~\citep{scipy}
and the $h_0^{95\%}$ value is
inferred along with its error bar.
As the detection statistic is
a Bayes factor $\BtSG$ marginalized over start times and durations \citep{1104.1704},
the upper limits at each injected $\tau$ value automatically include the full trials factor over the full allowed $\tau$ range of the search.
(I.e., the search is not optimized differently depending on $\tau$.)

\section{Details of the WPM analysis}
\label{app:FHwin}

For the \wpm search summarized in Section~\ref{sec:tcw-freqhough}, we provide here additional details on the setup and results,
as well as a summary of investigations into the three outlier candidates produced by the search.

\subsection{Details on WPM setup and results}

\begin{deluxetable}{lr}
\tablecaption{
 \label{tab:FH_setup}
 Main parameters of the \wpm search.
}
\tablehead{ \colhead{parameter} & \colhead{value}}
\startdata
$T_\mathrm{start}$ [MJD] & 60429.8466  \\
$\Ttot$ [days] & 101.7 \\
$T_\mathrm{ref}$ [MJD] & 60429.86975 \\
$[f_\mathrm{start},~f_\mathrm{stop}]$ [Hz] & [22.306, 22.506] Hz\\
$\fgw$ [Hz] & 22.365772235265929 \\
$\fdot$ [Hz\,s$^{-1}$] & $-3.128837145670767\times 10^{-11}$ \\
$\fddot$ [Hz\,s$^{-2}$] & $1.248409284183266\times 10^{-21}$ \\
$\fdddot$ [Hz\,s$^{-3}$] & $2.566286301492900\times 10^{-28}$ \\
\enddata
\tablecomments{
 $T_\mathrm{start}$ is the time of the first sample of the analyzed dataset, 2000 seconds before the glitch.
 $T_\mathrm{ref}$ is the reference time for Vela's position and rotational parameters, and corresponds to the glitch time. 
 $[f_\mathrm{start},~f_\mathrm{stop}]$ defines the frequency interval over which the search is run.
 The values of the frequency $\fgw$ and the first order spin-down, as used for heterodyning, include the variation due to the glitch, $2\Delta f_\mathrm{rot},~2\Delta \dot{f}_\mathrm{rot}$ respectively, see Table~\ref{tab:Vela_glitch}.  
}
\end{deluxetable}

The \wpm search uses the time and frequency parameters reported in Table~\ref{tab:FH_setup}.
The segment duration is chosen so that the residual, uncorrected, spin-down due to a \ns deformation induced by the glitch would produce a frequency variation confined within a single frequency bin.
Moreover, the Doppler modulation correction is robust with respect to a frequency mismatch (i.e. a difference between the frequency used in the heterodyne step, $\fgw$ in Table~\ref{tab:FH_setup}, and the actual signal frequency) as large as $\simeq0.2$\,Hz, larger than half the analyzed band.

A relevant issue in the search setup concerns the choice of the optimal observation window and segment duration.
As a general rule, it is not convenient to consider an observation window which extends to times at which the signal amplitude is significantly decreased, as this would correspond to diluting the signal peaks in the noise.

We have studied the optimal window duration (considering a range from 0.5 days to $\Ttot$, which is the total amount of data at our disposal) and segment duration (among four possible values of $\Tseg$ equal to 48000, 68000, 96000 and 172332 seconds) assuming both a signal with a single exponential decay, with various decay times, and a generalization of the model described by \citet{2204.12869}, describing the amplitude evolution by a combination of three decaying exponentials, with decay times derived from the frequency evolution observed in the radio.
Given one of these signal models, for each value of the observing window and of the segment duration, we have injected simulated signals of different amplitudes into O4b data and run the analyses, computing for each case the maximum $CR$.
For each signal model, the values $\Tobs\best$ and $\Tseg\best$ that maximize the $CR$ are our optimal choice.

\begin{figure}
    \centering
    \includegraphics{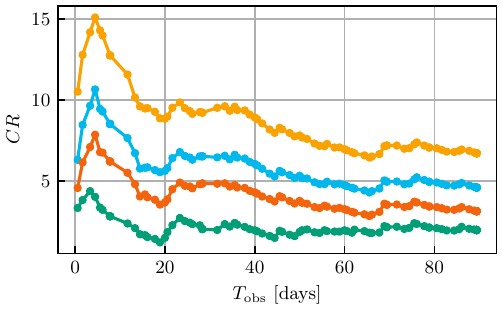}
    \caption{$CR$ as a function of the observing window for the multi-$\tau$ signal model and using the optimal segment duration $\Tseg\best=96000$ seconds.
    The plot has been obtained injecting simulated signals into L1 data with four different signal amplitudes (going from the bottom to the top curve: $[2.1,~4.2,~6.3,~8.4]\times 10^{-24}$).
    The ``optimal'' observing window corresponds to the maximum of the $CR$, which is nearly independent of the signal amplitude, taking place at $\Tobs\best\simeq 5.5$ days.}
    \label{fig:topt}
\end{figure}

As an example, in Figure~\ref{fig:topt} we show the $CR$ as a function of the observing window for the multi-$\tau$
signal model, and for which the optimal segment duration is $\Tseg\best=96000$\,s, for different signal amplitudes ($[2.1,~4.2,~6.3,~8.4]\times 10^{-24}$) injected into L1 O4b data.
The maximum of the $CR$ is at $\Tobs\best\simeq 5.5$ days.

\begin{figure}
    \centering
    \includegraphics[width=0.495\linewidth]{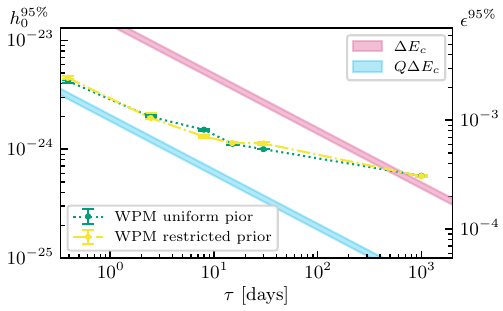}
    \caption{Upper limits ($95\%$ C.L.) from the \wpm search in terms of strain amplitude and \ns ellipticity immediately after the glitch, as a function of the signal decay time $\tau$.
    The two lines correspond to the upper limits for, respectively, uniform and constrained priors on polarization parameters.
    See Section~\ref{sec:tcw-cwinpy} for a discussion on the constrained parameters.
    For each $\tau$ the search is run on a time window $\Tobs\best$ and using segments of duration $\Tseg\best$ as shown in Table~\ref{tab:FH_summary}.
    \label{fig:cw_uls_wpm}
    }
\end{figure}

The full results of this optimization process are shown in columns 2 and 3 of Table~\ref{tab:FH_summary},
together with the numerical values of the final upper limits derived from the search.
The full set of upper limits is also plotted in Figure~\ref{fig:cw_uls_wpm}. 

\begin{deluxetable}{|c|c|c|c|c|}
\tablecaption{
 \label{tab:FH_summary}
 Summary of the setup choices for the \wpm search and $95\%$ C.L. upper limits obtained for each search configuration.
}
\tablehead{ \colhead{$\tau$ [d]} & \colhead{$\Tobs\best$ [d]} & \colhead{$\Tseg\best$ [s]} & \colhead{$h^\mathrm{95\%,restr}_\mathrm{0}$} & \colhead{$h^\mathrm{95\%,unif}_\mathrm{0}$}}
\startdata
$15.11,2.46,0.39$  & 5.5 & 96000 & $4.7\times 10^{-24}$ & $4.5\times 10^{-24}$\\
$0.39$ & 1.1 & 68000 &  $4.5\times 10^{-24}$ & $4.2\times 10^{-24}$\\
$2.46$ & 5.0 & 96000 & $1.9\times 10^{-24}$ & $2.0\times 10^{-24}$\\
$8$ & 6.0 & 96000 & $1.3 \times 10^{-24}$ & $1.5 \times 10^{-24}$\\
$15.11$ & 5.5 & 96000 & $1.1\times 10^{-24}$ & $1.1\times 10^{-24}$\\
$30$ & 25 & 172332 & $1.1\times 10^{-24}$ & $1.0\times 10^{-24}$\\
$1000$ & 101.7 & 172332 & $5.6\times 10^{-25}$ & $5.7\times 10^{-25}$\\
\enddata
\tablecomments{
 The first row corresponds to a search for a signal with amplitude varying in time according to a combination of three exponentials, with decay times inferred from the frequency evolution seen in radio observations.
 The other rows assume a signal amplitude described by a single exponential decay.
 The last row describes a signal with a nearly constant amplitude.
 $\Tobs\best$ and $\Tseg\best$ are, respectively, the optimal observation window and segment duration for the assumed signal model.
 These values depend not only on the signal model but also on the presence and position of gaps in the data.
 The penultimate column gives the upper limit for restricted priors on polarization parameters, and the last column for unrestricted uniform priors.
}
\end{deluxetable}

Upper limits are computed by simulated signal injections into the data, taking the amplitude such that $95\%$ of the injected signals have a $CR$ larger than the highest $CR$ found in the analysis with the same setup, taking into account the specific 0.02\,Hz sub-band in which each injection has been done.

\subsection{Follow-up of outliers}
\label{app:wpm_fu}

In this section, we discuss the follow-up of the three outliers found in the \wpm search.

The second and third outliers are likely due to an instrumental noise line.
Indeed, by re-running the analysis \textit{without} Doppler correction, their frequency shifts to 22.5\,Hz and the corresponding $CR$ increases significantly (by almost a factor of two), as expected for a detector disturbance.

The first outlier, which is more difficult to discard, 
has frequency 22.45538\,Hz, spin-down $-3.108294\times10^{-11}$\,Hz/s (at reference GPS time 1396742418), $CR=6.65$ and has been found in the search configuration with $\tau=8$\,days, $\Tobs=6$\,days and $T_\mathrm{seg}=96000$ seconds.
Most of the follow-up tests we have done have not shown a clear incompatibility with an astrophysical signal. 

As a first veto, we have again re-run the analysis switching off the Doppler modulation correction.
This results in a decrease in $CR$ from 6.65 to -0.96, which is compatible with what we expect for a real signal.

A second test consisted in running a semi-coherent search, using the semi-coherent 5-vector method \citep{DAntonio:2023jxm}, with segment duration of 2, 3 and 6 days.
Actually, these analyses produces three nearby candidates, with very similar $CR$, compatible with the original one, which are all considered in the next steps.  
The candidates remain significant when a segment duration of 2 days is used.
For an astrophysical signal we expect the $CR$ to increase with longer segment duration.
This does \textit{not} happen when passing from 2 to 3 days (the $CR$ remains nearly unchanged), while it happens when 6 days are used (going from $\sim$ 7.5 at 2 days to $\sim$ 10.7 at 6 days). 

We then repeated the analysis with a single data segment of 6 days shifting the starting time both backward and forward (-3.0, -1.5, -0.5, +0.5, +1.5, +3 days with respect to the signal reference starting time).
The resulting $CR$s have been compared with what we would expect for a real signal which starts around the glitch time, both assuming an exponential decay with $\tau=8$ days over a time window of 6 days and a constant amplitude signal of duration 5 or 7 days.
We find that the $CR$ steadily decreases both when the observing window is shifted backward and forward, becoming fully compatible with noise for a shift of $\pm 3$ days. 
Overall, this test seems to indicate that there is a feature in the data which switches on around the glitch time and switches off after 5 to 7 days.

\begin{deluxetable}{|c|c|c|c|c|c|c|c|}
\tablecaption{
 \label{tab:wpm_cand_5vect}
 Main parameters of the \wpm outliers after coherent 5-vector follow-up.
}
\tablehead{ \colhead{Outlier} & \colhead{frequency [Hz]} & \colhead{spin-down [Hz/s]} & \colhead{$h_0$} & \colhead{$\eta$} & \colhead{$\psi$ [deg]} &\colhead{$CR$} & \colhead{$\mathcal{C}$} }
\startdata
1 & 22.455332 & $-3.316\times 10^{-11}$ & $5.62\times 10^{-25}$ & -0.455 & 9.79 & 10.6 & 0.81\\
2 & 22.455333 & $-3.129\times 10^{-11}$ & $7.72\times 10^{-25}$ & 0.037 & 6.78 & 10.7 & 0.91\\
3 & 22.455334 & $-3.129\times 10^{-11}$ & $4.64\times 10^{-25}$ & -0.20 & 9.80 & 10.7 & 0.54\\
\enddata
\tablecomments{Parameter estimation, with the fully-coherent 5-vector method, for the three candidates associated to the first outlier from the main search.
Rotational parameters are referred to MJD time 60429.8466 (GPS 1398457164.24).}
\end{deluxetable}

Finally, we have used the fully-coherent 5-vector method \citep{Astone:2010zz} to estimate the parameters of the potential signal associated to the candidates from the semi-coherent 5-vector step, using the frequency and spin-down values found in the $\Tseg=6$\,days analysis. 
Estimated parameters for the three candidates we have analyzed are shown in Table~\ref{tab:wpm_cand_5vect}.
The last column, in particular, is an estimation of the so-called 5-vector ``coherence'', a measure (between 0 and 1) of how much the data resembles the best matching template.
The highest value we get, $\mathcal{C}=0.91$ for the second candidate, corresponds to a false-alarm probability (in Gaussian noise) of about 0.012, after taking the trial factor into account.

\begin{figure}
    \centering
      \includegraphics{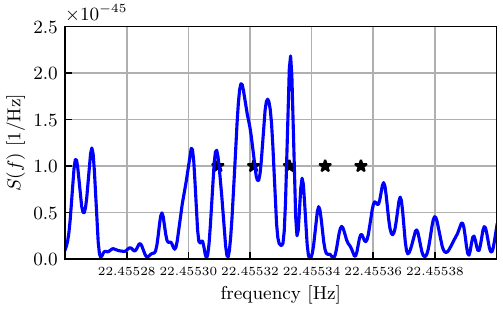} \\
      \includegraphics{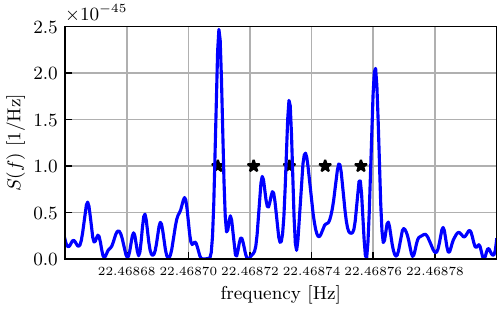}
      \includegraphics{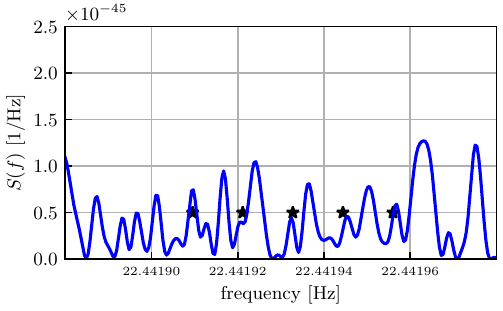}
    \caption{Top plot: power spectrum $S(f)$ of the LIGO data, after correction of the Doppler effect and spin-down of the second candidate in Table~\ref{tab:wpm_cand_5vect}.
    The two other plots show the corrected power spectrum after the injection of a simulated signal with the same parameters as the second candidate in Table~\ref{tab:wpm_cand_5vect} but with frequency shifted by $+13.4$\,mHz (bottom left) and $-13.4$\,mHz (bottom right).
    The asterisks denote the position of the five expected peaks produced by the sidereal modulation.}
    \label{fig:wpm_cand_spectrum}
\end{figure}

We have then inspected the power spectrum of the data, after Doppler and spin-down correction: in presence of a sufficiently strong astrophysical signal, we expect to see the typical 5-vector signature, due to the signal's sidereal modulation.
The spectrum after correcting the data for the parameters of the second candidate in Table~\ref{tab:wpm_cand_5vect} (that with the highest \textit{coherence}) is shown in the top plot of Figure~\ref{fig:wpm_cand_spectrum}.
The five asterisks denote the expected frequency of the five peaks (not the amplitude).
We can see that at least two, out of five, peaks have a rather clear counterpart in the data.
As a cross-check, we have computed the spectrum after injecting a simulated signal with the same parameters as those estimated for the second candidate, except for the frequency that has been shifted by some random amount.
As an example in Figure~\ref{fig:wpm_cand_spectrum} we show the spectra for shifts of $\pm13.4$\,mHz.
At the higher frequency, the spectrum looks slightly better when compared to the expected peak position, with two peaks clearly corresponding to features in the data (the first and the third peak, as for the real analysis candidate) and, possibly, also the fourth.
At the lower frequency, on the other hand, the situation is much less clear, with only one convincing correspondence among the expected peak frequencies and the features in the data.

Overall, our tests for this outlier have been inconclusive.
Our conclusion is that, even if a \gw signal was really present in the data, it is very marginal, and a detection cannot be claimed with sufficient confidence. 

\section{Details of the HMM analysis}
\label{app:hmm}

Here we provide additional details concerning the \hmm search. 
The template parameters, search band, and coherence times are summarized in Table~\ref{tab:HMM_setup}. 

\begin{deluxetable}{lc}
\tablecaption{
 \label{tab:HMM_setup}
 Search setup for the \hmm search.
}
\tablehead{ \colhead{parameter} & \colhead{value}}
\startdata
$T_\mathrm{start}$ [GPS] & $1398458994$ \\
$\Tref$ [GPS] & $1398459095.043$ \\
$\Tobs$ [days] & $90$ \\
$\Tcoh$ [days] & $0.375$, $2.4374$, $15.1042$ \\
$\frot$ [Hz] & $11.1828595$ \\
$\fdotrot$ [Hz/s] & $-1.5540586\times 10^{-11}$ \\
$\fddotrot$ [Hz/s$^2$] & $6.242\times 10^{-22}$ \\
$\fdddotrot$ [Hz/s$^3$] & $1.28\times 10^{-28}$ \\
$\Delta \frot$ [Hz] & $0.05$ \\
$N_T$ & $240$, $36$, $5$ \\
\enddata
\tablecomments{
The GPS start time is rounded to the last \sft timestamp before the glitch arrival time at the detectors.
The coherence times are similarly rounded down to the nearest multiple of $1800$\,s.
The reference time is the glitch arrival time at the solar system barycenter, and the spin frequencies are extrapolated to this time.
The frequency, bandwidth, and frequency derivative correspond to the Vela spin frequency; the values at twice the spin frequency are found by multiplying these quantities by two.
}
\end{deluxetable}

As discussed in Section~\ref{sec:tcw-hmm}, empirical detection thresholds are set using one of two methods.
In the first method, we perform the search on $10^4$ realizations of synthetic Gaussian noise data to accumulate log-likelihood samples $\bar{\mathcal{L}}_i$, keeping the same search parameters given in Table~\ref{tab:HMM_setup}.
We set thresholds based on the Viterbi log-likelihood $\bar{\mathcal{L}}_\mathrm{th}$ at which a desired fraction $\alpha$ of the samples derived from Gaussian noise simulations are above the threshold,
$\bar{\mathcal{L}}_i>\bar{\mathcal{L}}_\mathrm{th}$.
In the second method, we set thresholds in exactly the same way, but instead the log-likelihood samples are derived from searching real detector data at $10^4$ off-target sky positions.
We sample off-target sky positions by shifting the search from the true Vela position by a random angle between $\pm (5^\circ\text{--}15^\circ)$ in RA and $\pm 5^\circ$ in Dec.
In both methods, we set the false-alarm probability at $\alpha=0.01/3$ for each choice of $\Tcoh$, such that the overall rate of false alarms, accounting for the trials factor from three different coherence times, is $1\%$.
The resulting thresholds for each choice of $\Tcoh$ and sampling method are shown in Table~\ref{tab:HMM_results}, where we also list the maximum statistics obtained from the single and dual harmonic \hmm searches.

\begin{deluxetable}{lccc}
\tablecaption{
 \label{tab:HMM_results}
 Detection statistic thresholds for the \hmm search.
}
\tablehead{ \colhead{Statistic} & \colhead{$\Tcoh=0.375\,\,\mathrm{days}$} & \colhead{$2.4374\,\,\mathrm{days}$} & \colhead{$15.1042\,\,\mathrm{days}$} }
\startdata
$\bar{\mathcal{L}}_\mathrm{th,G}^\mathrm{2F}$ & $7.05$ & $9.33$ & $17.14$ \\
$\bar{\mathcal{L}}_\mathrm{th,OT}^\mathrm{2F}$ & $6.99$ & $9.32$ & $17.06$ \\
$\bar{\mathcal{L}}_\mathrm{max}^\mathrm{2F}$ & $6.85$ & $8.43$ & $14.44$ \\
\hline
$\bar{\mathcal{L}}_\mathrm{th,G}^\mathrm{1F+2F}$ & $11.86$ & $14.77$ & $24.43$ \\
$\bar{\mathcal{L}}_\mathrm{th,OT}^\mathrm{1F+2F}$ & $11.5$ & $14.38$ & $24.09$ \\
$\bar{\mathcal{L}}_\mathrm{max}^\mathrm{1F+2F}$ & $11.46$ & $13.56$ & $19.59$ \\
\enddata
\tablecomments{All log-likelihood thresholds are quoted at the level of $1\%$ false-alarm probability, including trials factors.}
\end{deluxetable}

\begin{figure}
\centering
\includegraphics[width=0.495\linewidth]{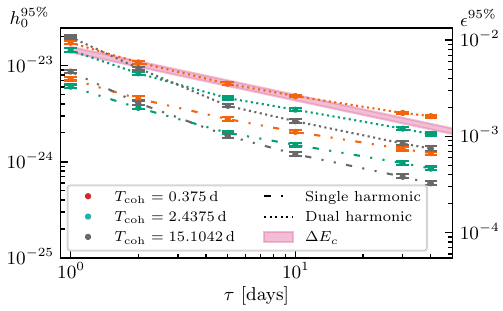}
\caption{ 
\label{fig:h0_uls_hmm}
Upper limits on long-transient \gws from Vela inferred by the single (dash-dotted lines) and dual-harmonic (dotted lines) \hmm analyses as a function of the exponential signal decay time parameter $\tau$.
Upper limits are expressed in terms of strain amplitude $h_0^{95\%}$ (left axis) and \ns ellipticity $\epsilon^{95\%}$ (right axis).
The magenta shaded comparison band labeled $\Delta\Echar$ corresponds to Equation~\eqref{eq:tcw_indirect_UL_PGM}
with $Q=1$.
}
\end{figure}

In simulating the injections used to set upper limits, the relative amplitudes of each harmonic depend on the angle, $\theta$, between the \ns axis of rotation and the principal moment of inertia, where $\theta=\pi/2$ corresponds to a \gw emission at $\fgw=2\frot$ only \citep[e.g.,][]{gr-qc/9804014}.
We therefore fix $\theta=\pi/2$ for the single harmonic injections and $\theta=\pi/4$ for the dual harmonic injections, where the latter corresponds to roughly equal strain amplitudes at $\frot$ and $2\frot$.
Figure~\ref{fig:h0_uls_hmm} shows upper limits on the strain amplitude and ellipticity, respectively, inferred by the \hmm analysis. 
Different curves correspond to different choices of the coherence time and whether the signal model is assumed to have one component at twice the Vela spin frequency or at both once and twice the spin frequency.
Unfortunately, because of the large disparity in the detector noise floors between $11$\,Hz and $22$\,Hz, the band at $\fgw=\frot$ mostly contributes noise for any reasonable range of $h_0$.
Hence, our upper limits that assume dual harmonic emission are consistently worse by a factor of $\sim 2.3$.

\section{Details of the joint NS properties inference}
\label{app:joint}

To sample the parameter space for joint inference under the assumption of the Vela pulsar emitting \gws both from an f-mode and a transient mountain, we use the following priors.
The \ns mass prior is uniform in $[0.9 \, M_{\odot}, 2.6 \, M_{\odot}]$.
The \ns radius prior is uniform in $[8 \, \mathrm{km}, 16 \, \mathrm{km}]$.
Constraints are enforced such that the mass and radius obey causality and the \ns does not collapse to a black hole (see black contour in Figure~\ref{fig:MR_corner_plot_variable_fmode_CW}).
The prior for the glitch-induced spin change $\Delta \frot/\frot$ is a Gaussian distribution centered at $2.377\times10^{-6}$ with $\sigma = 4.47\times10^{-11}$.
We choose the inclination angle $\iota$ prior to be a double-peaked Gaussian distribution with peaks at 1.1048 and 2.0368 radians and $\sigma=10.5$\,mrad in order to match the astrophysical prior used in Section~\ref{sec:tcw-cwinpy} and based on \citet{2008ApJ...673..411N}.
The distance $d$ prior is a Gaussian centered on $\distvela \, \mathrm{pc}$ with $\sigma=9.5 \, \mathrm{pc}$,
matching \citet{2003ApJ...596.1137D}.
The mountain decay time $\tau_{\mathrm{CW}}$ prior is uniform from $1$ hour to $120$ days,
matching the search durations covered in Section~\ref{sec:tcw}.
The prior for the mountain energy scaling factor $\digamma_{\mathrm{CW}}$ is log-uniform from $10^{-4}$ to $1.0$
(none of the searches would be sensitive below the cutoff)
and the prior for the f-mode energy scaling factor $\digamma_{\mathrm{fmode}}$ is uniform between $0$ and $1$ (Figure~\ref{fig:MR_corner_plot_variable_fmode_CW}) or fixed to $0.8$ (Figure~\ref{fig:mass_vs_radius_fixed_fmode_variable_CW}).
For the model-specific results reported in Figure~\ref{fig:mass_vs_radius_fixed_fmode_variable_CW},
we also set the prior on the duration parameter $\tau_{\mathrm{CW}}$ of CW-like signals to a Gaussian distribution peaked at twice the first decay time $\tau_d^1 = 15.1\, \mathrm{days}$ of the radio frequency evolution, with $\sigma=0.2\,\mathrm{days}$, to match the \citet{2007.05893} model and Table~\ref{tab:Vela_glitch}.

We then combine the non-detection results from one representative pipeline each for the burst-type search for f-modes and the \cw-like search for long transients from post-glitch mountain formation.
For the f-mode, we use the ringdown injection curves generated by \xpipe for different frequencies and damping times.
These curves are linearly interpolated and subtracted from unity to give the probability that a waveform generated by a particular set of parameters would be detected by \xpipe.
For the mountain model, we use the posterior distribution on amplitude $h_0$ and \cw damping time $\tau_{\mathrm{CW}}$ produced by \cwinpy for an exponential decay profile
and constrained orientation, as described in appendix~\ref{app:cwinpy}.
This likelihood distribution is interpreted as the probability that a given parameter would generate a waveform consistent with background noise.

For each sample, we compute the probability of non-detection using the \xpipe-informed likelihood function and the probability of consistency with background using the \cwinpy-informed likelihood function.
We then re-weigh the prior samples with the product of these likelihoods according to Bayes' Theorem to give a posterior distribution on the sampled parameters consistent with search results.
This re-weighting is done via rejection sampling, where the product of the two likelihoods is used as the probability that a given sample will not be rejected.
Samples with higher likelihood are thus more likely to be preserved, and samples with lower likelihoods are more likely to be rejected.

Further details of the implementation are described in \citet{ball_joint_implications}.

\bibliography{o4vela}{}
\bibliographystyle{aasjournal}


\iftoggle{endauthorlist}{
\expandafter\gdef\csname
currCollabName\the\allentries\endcsname{}
\allauthors
}


\end{document}

%% file: macros.tex
\newcommand{\lvkfull}{The LIGO Scientific Collaboration, the Virgo Collaboration, and the KAGRA Collaboration}
\newcommand{\lvkaff}{(\lvkfull)}

\newcommand{\dcc}{P2500086}
\newcommand{\dccurl}{https://dcc.ligo.org/\dcc/public}
\newcommand{\dcclink}{\href{\dccurl}{LIGO-\dcc}\xspace}

\acrodef{lvk}[LVK]{LIGO--Virgo--KAGRA}
\newcommand{\lvk}{\ac{lvk}\xspace}
\acrodef{gw}[GW]{gravitational-wave}
\acrodefplural{gw}[GWs]{gravitational waves}
\newcommand{\gw}{\ac{gw}\xspace}
\newcommand{\gws}{\acp{gw}\xspace}
\acrodef{cw}[CW]{continuous wave}
\acrodefplural{cw}[CWs]{continuous waves} 
\newcommand{\cw}{\ac{cw}\xspace}
\newcommand{\cws}{\acp{cw}\xspace}
\acrodef{ns}[NS]{neutron star}
\acrodefplural{nss}[NSs]{neutron stars}
\newcommand{\ns}{\ac{ns}\xspace}
\newcommand{\nss}{\acp{ns}\xspace}
\acrodef{elmag}[EM]{electromagnetic}
\newcommand{\elmag}{\ac{elmag}\xspace}
\acrodef{snr}[S/N]{signal-to-noise ratio}
\acrodefplural{snrs}[S/N]{signal-to-noise ratios}

\acrodef{sft}[SFT]{short Fourier transform}
\acrodefplural{sft}[SFTs]{short Fourier transforms}
\newcommand{\sft}{\ac{sft}\xspace}
\newcommand{\sfts}{\acp{sft}\xspace}
\acrodef{fft}[FFT]{fast Fourier transform}
\acrodefplural{fft}[FFTs]{fast Fourier transforms}

\acrodef{sfdb}[SFDB]{Short Fourier Database}
\acrodefplural{sfdb}[SFDBs]{Short Fourier Databases}

\acrodef{bsd}[BSD]{band-sampled data}
\acrodefplural{bsd}[BSDs]{band-sampled data}
\newcommand{\bsd}{\ac{bsd}\xspace}
\newcommand{\bsds}{\acp{bsd}\xspace}
\acrodef{psd}[PSD]{power spectral density}
\acrodefplural{psd}[PSDs]{power spectral densities}
\newcommand{\psd}{\ac{psd}\xspace}

\acrodef{asd}[ASD]{amplitude spectral density}
\acrodefplural{asd}[ASDs]{amplitude spectral densities}
\newcommand{\asd}{\ac{asd}\xspace}
\newcommand{\asds}{\acp{asd}\xspace}
\acrodef{ul}[UL]{upper limit}
\acrodefplural{uls}[ULs]{upper limits}

\acrodef{ohthree}[O3]{third observing run}

\acrodef{ohfour}[O4]{fourth observing run}
\newcommand{\ohfour}{\ac{ohfour}\xspace}
\newcommand{\cwb}{\texttt{cWB}\xspace}
\newcommand{\pystampas}{\texttt{PySTAMPAS}\xspace}
\newcommand{\xpipe}{\texttt{X-pipeline}\xspace}
\acrodef{hmm}[HMM]{hidden Markov model\xspace}
\newcommand{\hmm}{\ac{hmm}\xspace}
\acrodef{wpm}[WPM]{weighted peakmaps\xspace}
\newcommand{\wpm}{\ac{wpm}\xspace}
\newcommand{\cwinpy}{\texttt{CWInPy}\xspace}
\newcommand{\tcwip}{\texttt{tCWip}\xspace}
\newcommand{\lalsuite}{\texttt{LALSuite}\xspace}
\newcommand{\pyfstat}{\texttt{PyFstat}\xspace}
\acrodef{iar}[IAR]{Argentine Institute of Radio Astronomy}
\newcommand{\IAR}{\ac{iar}\xspace}
\acrodef{mpro}[MPRO]{Mount Pleasant Radio Observatory}
\newcommand{\MPRO}{\ac{mpro}\xspace}
\acrodef{toa}[TOA]{Time of Arrival}
\acrodefplural{toas}[TOAs]{Times of Arrival}

\newcommand{\toas}{\acp{toa}\xspace}

\acrodef{mjd}[MJD]{Modified Julian Date}
\newcommand{\mjd}{\ac{mjd}\xspace}

\newcommand{\glitch}{\mathrm{gl}}
\newcommand{\Tgl}{T_\glitch} 
\newcommand{\Egw}{E_\mathrm{gw}} 
\newcommand{\frot}{f_{\mathrm{rot}}} 
\newcommand{\fdotrot}{\dot{f}_\mathrm{rot}}
\newcommand{\fddotrot}{\ddot{f}_\mathrm{rot}}
\newcommand{\fdddotrot}{\dddot{f}_\mathrm{rot}}
\newcommand{\fgw}{f_{\mathrm{gw}}} 
\newcommand{\fdot}{{\dot{f}_{\mathrm{gw}}}} 
\newcommand{\fddot}{\ddot{f}_{\mathrm{gw}}} 
\newcommand{\fdddot}{\dddot{f}_{\mathrm{gw}}} 
\newcommand{\F}{\mathcal{F}} 
\newcommand{\Fstat}{$\F$-statistic\xspace}
\newcommand{\Tobs}{T_\mathrm{obs}} 
\newcommand{\Tcoh}{T_\mathrm{coh}} 
\newcommand{\Tseg}{T_\mathrm{seg}} 
\newcommand{\Ttot}{T_\mathrm{tot}} 
\newcommand{\best}{^\mathrm{best}} 

\newcommand{\Tref}{T_\mathrm{ref}} 
\newcommand{\Dop}{\lambda} 
\newcommand{\Ntemp}{N_{\Dop}} 
\newcommand{\BtSG}{\mathcal{B}_{tS/G}} 
\newcommand{\hrss}{h_{\mathrm{rss}}}
\newcommand{\Echar}{E_{\text{c}}} 

\newcommand{\distvela}{287} 
\newcommand{\distvelapm}{\ensuremath{\distvela_{-17}^{+19}}} 
\newcommand{\glitchdate}{29 April 2024\xspace}
\newcommand{\Qvalue}{0.017} 

%% file: LSC-Virgo-KAGRA-Authors-Feb-2025-aas.tex


\author[0000-0003-4786-2698]{A.~G.~Abac}
\affiliation{Max Planck Institute for Gravitational Physics (Albert Einstein Institute), D-14476 Potsdam, Germany}
\author{I.~Abouelfettouh}
\affiliation{LIGO Hanford Observatory, Richland, WA 99352, USA}
\author{F.~Acernese}
\affiliation{Dipartimento di Farmacia, Universit\`a di Salerno, I-84084 Fisciano, Salerno, Italy}
\affiliation{INFN, Sezione di Napoli, I-80126 Napoli, Italy}
\author[0000-0002-8648-0767]{K.~Ackley}
\affiliation{University of Warwick, Coventry CV4 7AL, United Kingdom}
\author[0000-0001-5525-6255]{C.~Adamcewicz}
\affiliation{OzGrav, School of Physics \& Astronomy, Monash University, Clayton 3800, Victoria, Australia}
\author[0009-0004-2101-5428]{S.~Adhicary}
\affiliation{The Pennsylvania State University, University Park, PA 16802, USA}
\author{D.~Adhikari}
\affiliation{Max Planck Institute for Gravitational Physics (Albert Einstein Institute), D-30167 Hannover, Germany}
\affiliation{Leibniz Universit\"{a}t Hannover, D-30167 Hannover, Germany}
\author[0000-0002-4559-8427]{N.~Adhikari}
\affiliation{University of Wisconsin-Milwaukee, Milwaukee, WI 53201, USA}
\author[0000-0002-5731-5076]{R.~X.~Adhikari}
\affiliation{LIGO Laboratory, California Institute of Technology, Pasadena, CA 91125, USA}
\author{V.~K.~Adkins}
\affiliation{Louisiana State University, Baton Rouge, LA 70803, USA}
% ...




%% file: em_authors.tex
\author[0000-0001-8691-8039]{Jim~Palfreyman}
\affiliation{School of Natural Sciences, University of Tasmania, Hobart, Australia}

\author[0000-0003-4027-4826]{S.B.~Araujo Furlan}
\affiliation{Instituto de Astronom\'{\i}a Te\'orica y Experimental, CONICET-UNC, Laprida 854, X5000BGR - C\'ordoba, Argentina}
\affiliation{Facultad de Matem\'atica, Astronom\'{\i}a, F\'{\i}sica y Computaci\'on, UNC. Av. Medina Allende s/n , Ciudad Universitaria, CP:X5000HUA - C\'ordoba, Argentina}

\author[0000-0002-5761-2417]{S.~del Palacio}
\affiliation{Instituto Argentino de Radioastronom\'{\i}a (CCT La Plata, CONICET; CICPBA; UNLP), C.C.5, (1894) Villa Elisa, Buenos Aires, Argentina}
\affiliation{Department of Space, Earth and Environment, Chalmers University of Technology, SE-412 96 Gothenburg, Sweden}

\author[0000-0003-1282-3031]{G.~Gancio}
\affiliation{Instituto Argentino de Radioastronom\'{\i}a (CCT La Plata, CONICET; CICPBA; UNLP), C.C.5, (1894) Villa Elisa, Buenos Aires, Argentina}

\author[0000-0001-9072-4069]{F.~Garc\'{i}a}
\affiliation{Instituto Argentino de Radioastronom\'{\i}a (CCT La Plata, CONICET; CICPBA; UNLP), C.C.5, (1894) Villa Elisa, Buenos Aires, Argentina}
\affiliation{Facultad de Ciencias Astron\'omicas y Geof\'{\i}sicas, Universidad Nacional de La Plata, Paseo del Bosque, B1900FWA La Plata, Argentina}

\author[0000-0002-5260-1807]{G.~E.~Romero}
\affiliation{Instituto Argentino de Radioastronom\'{\i}a (CCT La Plata, CONICET; CICPBA; UNLP), C.C.5, (1894) Villa Elisa, Buenos Aires, Argentina}
\affiliation{Facultad de Ciencias Astron\'omicas y Geof\'{\i}sicas, Universidad Nacional de La Plata, Paseo del Bosque, B1900FWA La Plata, Argentina}

\author[0009-0009-5593-367X]{E.~Zubieta}
\affiliation{Instituto Argentino de Radioastronom\'{\i}a (CCT La Plata, CONICET; CICPBA; UNLP), C.C.5, (1894) Villa Elisa, Buenos Aires, Argentina}
\affiliation{Facultad de Ciencias Astron\'omicas y Geof\'{\i}sicas, Universidad Nacional de La Plata, Paseo del Bosque, B1900FWA La Plata, Argentina}

%% file: P2000488_v28.tex
%
This material is based upon work supported by NSF's LIGO Laboratory, which is a
major facility fully funded by the National Science Foundation.
The authors also gratefully acknowledge the support of
the Science and Technology Facilities Council (STFC) of the
United Kingdom, the Max-Planck-Society (MPS), and the State of
Niedersachsen/Germany for support of the construction of Advanced LIGO 
and construction and operation of the GEO\,600 detector. 
Additional support for Advanced LIGO was provided by the Australian Research Council.
The authors gratefully acknowledge the Italian Istituto Nazionale di Fisica Nucleare (INFN),  
the French Centre National de la Recherche Scientifique (CNRS) and
the Netherlands Organization for Scientific Research (NWO)
for the construction and operation of the Virgo detector
and the creation and support  of the EGO consortium. 
The authors also gratefully acknowledge research support from these agencies as well as by 
the Council of Scientific and Industrial Research of India, 
the Department of Science and Technology, India,
the Science \& Engineering Research Board (SERB), India,
the Ministry of Human Resource Development, India,
the Spanish Agencia Estatal de Investigaci\'on (AEI),
the Spanish Ministerio de Ciencia, Innovaci\'on y Universidades,
the European Union NextGenerationEU/PRTR (PRTR-C17.I1),
the ICSC - CentroNazionale di Ricerca in High Performance Computing, Big Data
and Quantum Computing, funded by the European Union NextGenerationEU,
the Comunitat Auton\`oma de les Illes Balears through the Conselleria d'Educaci\'o i Universitats,
the Conselleria d'Innovaci\'o, Universitats, Ci\`encia i Societat Digital de la Generalitat Valenciana and
the CERCA Programme Generalitat de Catalunya, Spain,
the Polish National Agency for Academic Exchange,
the National Science Centre of Poland and the European Union - European Regional
Development Fund;
the Foundation for Polish Science (FNP),
the Polish Ministry of Science and Higher Education,
the Swiss National Science Foundation (SNSF),
the Russian Science Foundation,
the European Commission,
the European Social Funds (ESF),
the European Regional Development Funds (ERDF),
the Royal Society, 
the Scottish Funding Council, 
the Scottish Universities Physics Alliance, 
the Hungarian Scientific Research Fund (OTKA),
the French Lyon Institute of Origins (LIO),
the Belgian Fonds de la Recherche Scientifique (FRS-FNRS), 
Actions de Recherche Concert\'ees (ARC) and
Fonds Wetenschappelijk Onderzoek - Vlaanderen (FWO), Belgium,
the Paris \^{I}le-de-France Region, 
the National Research, Development and Innovation Office of Hungary (NKFIH), 
the National Research Foundation of Korea,
the Natural Sciences and Engineering Research Council of Canada (NSERC),
the Canadian Foundation for Innovation (CFI),
the Brazilian Ministry of Science, Technology, and Innovations,
the International Center for Theoretical Physics South American Institute for Fundamental Research (ICTP-SAIFR), 
the Research Grants Council of Hong Kong,
the National Natural Science Foundation of China (NSFC),
the Israel Science Foundation (ISF),
the US-Israel Binational Science Fund (BSF),
the Leverhulme Trust, 
the Research Corporation,
the National Science and Technology Council (NSTC), Taiwan,
the United States Department of Energy,
and
the Kavli Foundation.
The authors gratefully acknowledge the support of the NSF, STFC, INFN and CNRS for provision of computational resources.
This work was supported by MEXT,
the JSPS Leading-edge Research Infrastructure Program,
JSPS Grant-in-Aid for Specially Promoted Research 26000005,
JSPS Grant-in-Aid for Scientific Research on Innovative Areas 2402: 24103006,
24103005, and 2905: JP17H06358, JP17H06361 and JP17H06364,
JSPS Core-to-Core Program A.\ Advanced Research Networks,
JSPS Grants-in-Aid for Scientific Research (S) 17H06133 and 20H05639,
JSPS Grant-in-Aid for Transformative Research Areas (A) 20A203: JP20H05854,
the joint research program of the Institute for Cosmic Ray Research,
University of Tokyo,
the National Research Foundation (NRF),
the Computing Infrastructure Project of the Global Science experimental Data hub
Center (GSDC) at KISTI,
the Korea Astronomy and Space Science Institute (KASI),
the Ministry of Science and ICT (MSIT) in Korea,
Academia Sinica (AS),
the AS Grid Center (ASGC) and the National Science and Technology Council (NSTC)
in Taiwan under grants including the Science Vanguard Research Program,
the Advanced Technology Center (ATC) of NAOJ,
and the Mechanical Engineering Center of KEK.
The activities at the Instituto Argentino de
Radioastronom\'{i}a (IAR) are supported by the national agency
CONICET, the Province of Buenos Aires agency (CIC),
and the National University of La Plata (UNLP).

Additional acknowledgements for support of individual authors may be found in the following document: \\
\texttt{https://dcc.ligo.org/LIGO-M2300033/public}.
For the purpose of open access, the authors have applied a Creative Commons Attribution (CC BY)
license to any Author Accepted Manuscript version arising.
We request that citations to this article use 'A. G. Abac {\it et al.} (LIGO-Virgo-KAGRA Collaboration), ...' or similar phrasing, depending on journal convention.